%% file: main.tex
\pdfoutput=1
\documentclass[11pt]{article}

\usepackage[final]{acl}

\usepackage{times}
\usepackage{latexsym}

\usepackage[T1]{fontenc}
\usepackage[utf8]{inputenc}

\usepackage{microtype}

\usepackage{inconsolata}

\usepackage{graphicx}

\usepackage{comment}

\usepackage{hyperref}
\usepackage{url}
\usepackage{multirow}
\usepackage{booktabs}
\usepackage{graphicx}
\usepackage{xspace}
\usepackage{makecell}
\usepackage{xcolor}
\usepackage{bbding}
\usepackage{makecell}
\usepackage{listings}
\definecolor{codebg}{rgb}{0.95, 0.95, 0.92}

\definecolor{deepred}{rgb}{0.698,0.133,0.133}
\definecolor{blue}{rgb}{0,0,1}

\DeclareUnicodeCharacter{FF1B}{ }
\usepackage{tikz}
\usepackage[edges]{forest}
\usepackage{pgfplots}
\usepackage{tabularx}
\usepackage{enumitem}
\usepackage{amsmath}
\usepackage[normalem]{ulem}
\usepackage{threeparttable}

\usepackage{amsmath}           % Math symbols
\usepackage{amssymb}           % Additional math symbols
\usepackage{booktabs}          % Professional tables (\toprule, \midrule, etc.)
\usepackage{enumitem}          % Customizable lists
\usepackage{xcolor}            % Color support (for listings)
\usepackage{listings}          % Code listings
\usepackage{algorithm}         % Algorithm environment
\usepackage{algorithmic}     % Algorithmicx pseudocode (or use \usepackage{algorithmic})
\usepackage{tabularx}          % Better table formatting
\usepackage{array}             % Extended table features
\usepackage{multirow}          % Multi-row cells (if needed)

\lstdefinelanguage{Python}{
    morekeywords={import,from,class,def,return,if,else,for,while,try,except},
    sensitive=true,
    morecomment=[l]{\#},
    morestring=[b]',
    morestring=[b]"
}

\lstdefinelanguage{Java}{
    morekeywords={public,class,void,import,package,return,if,else,new,@Test},
    sensitive=true,
    morecomment=[l]{//},
    morecomment=[s]{/*}{*/},
    morestring=[b]",
}

\lstdefinelanguage{JavaScript}{
    morekeywords={const,let,var,function,return,if,else,require,module,exports},
    sensitive=true,
    morecomment=[l]{//},
    morecomment=[s]{/*}{*/},
    morestring=[b]',
    morestring=[b]"
}

\usepackage[edges]{forest}
\usepackage[framemethod=tikz]{mdframed}
\usepackage{subcaption}
\definecolor{lgreen}{rgb}{0.89,0.94,0.85}
\definecolor{lred}{rgb}{0.98, 0.90, 0.84}
\definecolor{lyellow}{rgb}{1.00, 0.95, 0.80}
\definecolor{lblue}{rgb}{0.85, 0.89, 0.95}
\definecolor{hidden-draw}{RGB}{20,68,106}
\definecolor{hidden-pink}{RGB}{255,245,247}

\tikzset{%
    parent/.style =          {align=center,text width=0.7cm, rounded corners=2pt, line width=0.8mm, fill=white!0, draw=white!90},
    child/.style =           {align=center,text width=1.4cm,rounded corners=2pt, fill=blue!10,draw=blue!90,line width=0.3mm},
    T1/.style =           {align=center,text width=1.8cm,rounded corners=3pt, fill=lblue!100, draw=black,line width=0.2mm},   
    T1_end/.style =           {align=left, text width=5cm,rounded corners=5pt, fill=lblue!100,draw=blue!0,line width=0.3mm},
    T2/.style =           {align=center,text width=1.8cm,rounded corners=3pt, fill=lred!100, draw=black,line width=0.2mm},   
    T2_end/.style =           {align=left, text width=5cm,rounded corners=5pt, fill=lred!100,draw=blue!0,line width=0.3mm},
    T3/.style =           {align=center,text width=1.8cm,rounded corners=3pt, fill=lyellow!100, draw=black,line width=0.2mm},   
    T3_end/.style =           {align=left, text width=5cm,rounded corners=5pt, fill=lyellow!100,draw=blue!0,line width=0.3mm},
    T4/.style =           {align=center,text width=1.8cm,rounded corners=3pt, fill=lgreen!100, draw=black,line width=0.2mm},   
    T4_end/.style =           {align=left, text width=5cm,rounded corners=5pt, fill=lgreen!100,draw=blue!0,line width=0.3mm}
}

\mdfdefinestyle{dataset}{
	innertopmargin=0.1\baselineskip,
	innerbottommargin=0.1\baselineskip,
	roundcorner=5pt,
	nobreak,
	singleextra={%
	},
}

\newenvironment{dataset}[1][\unskip]{
	\newcommand{\promptVanillaTitle}{Python Project Example}
        \begin{figure}[t]
        \centering
        \begin{minipage}{0.7\linewidth}
	\begin{mdframed}[style=dataset, backgroundcolor=white]\small
}{
        \includegraphics[width=\linewidth]{./sections/figures/project_example3.pdf}
	\end{mdframed}
        \end{minipage}
        \caption{An example of MultiFileTest.}
        \label{fig: dataset}
        \end{figure}
}

\mdfdefinestyle{unittest}{
	innertopmargin=1.2\baselineskip,
	innerbottommargin=0.8\baselineskip,
	roundcorner=5pt,
	nobreak,
	singleextra={%
		\draw(P-|O)node[xshift=1em,anchor=west,fill=yellow!15,draw,rounded corners=5pt]{%
		  \promptVanillaTitle};
	},
}

\mdfdefinestyle{prompt}{
	innertopmargin=0.7\baselineskip,
	innerbottommargin=0.1\baselineskip,
	roundcorner=5pt,
	nobreak,
	singleextra={%
		\draw(P-|O)node[xshift=1em,anchor=west,fill=red!15,draw,rounded corners=5pt]{%
		 \scriptsize \promptVanillaTitle};
	},
}
\newenvironment{prompt}[1][\unskip]{
	\newcommand{\promptVanillaTitle}{Vanilla Prompt for Python}
        \begin{figure}[t]
        \centering
	\begin{mdframed}[style=prompt, backgroundcolor=white]\small
}{
	\end{mdframed}
        \caption{The prompt used to generate unit tests for Python projects. \textcolor{purple}{Purple indicates language-specific instruction.} \textcolor{blue}{Blue}, \textcolor{orange}{orange}, and \textcolor{red}{red} indicates instructions related to executability rate, correctness rate, and coverage rate, respectively.}
        \label{fig: prompt}
        \end{figure}
}
\newenvironment{self_fix_prompt}[1][\unskip]{
	\newcommand{\promptVanillaTitle}{Self-fixing Prompt for Python}
        \begin{figure}[t]
	\begin{mdframed}[style=prompt, backgroundcolor=white]\small
}{
	\end{mdframed}
        \caption{The prompt used for the LLM self-fixing scenario for Python projects.}
        \label{fig: self_fix_prompt}
        \end{figure}
}
\newenvironment{prompt_java}[1][\unskip]{
	\newcommand{\promptVanillaTitle}{Vanilla Prompt for Java}
        \begin{figure}[h]
        \centering
	\begin{mdframed}[style=prompt, backgroundcolor=white]\small
}{
	\end{mdframed}
        \caption{The prompt used to generate unit tests for Java projects. }
        \label{fig: prompt_java}
        \end{figure}
}
\newenvironment{prompt_js}[1][\unskip]{
	\newcommand{\promptVanillaTitle}{Vanilla Prompt for JavaScript}
        \begin{figure}[h]
        \centering
	\begin{mdframed}[style=prompt, backgroundcolor=white]\small
}{
	\end{mdframed}
        \caption{The prompt used to generate unit tests for JavaScript projects. }
        \label{fig: prompt_js}
        \end{figure}
}

\newenvironment{prompt_comment}[1][\unskip]{
	\newcommand{\promptVanillaTitle}{Prompt for Python with Comment Sign}
        \begin{figure}[h]
        \centering
	\begin{mdframed}[style=prompt, backgroundcolor=white]\small
}{
	\end{mdframed}
        \caption{The prompt used to generate unit tests for Python projects. }
        \label{fig: prompt_comment}
        \end{figure}
}

\mdfdefinestyle{errors}{
	innertopmargin=0.1\baselineskip,
	innerbottommargin=0.1\baselineskip,
	roundcorner=5pt,
	nobreak,
	singleextra={%
	},
}

\newenvironment{compilation_errors}[1][\unskip]{
	\newcommand{\promptVanillaTitle}{Executability Error Example}
        \begin{figure}[t]
        \centering
        \begin{minipage}{0.9\linewidth}
	\begin{mdframed}[style=errors, backgroundcolor=white]\small
}{
        \includegraphics[width=\linewidth]{./sections/figures/compilation_error3.pdf}
	\end{mdframed}
        \end{minipage}
        \caption{Executability error (tests don't run).}
        \label{fig: compilation_errors}
        \end{figure}
}

\newenvironment{cascade_errors}[1][\unskip]{
	\newcommand{\promptVanillaTitle}{Cascade Error Example}
        \begin{figure}[t]
        \centering
        \begin{minipage}{0.9\linewidth}
	\begin{mdframed}[style=errors, backgroundcolor=white]\small
}{
        \includegraphics[width=\linewidth]{./sections/figures/cascade_error3.pdf}
	\end{mdframed}
        \end{minipage}
        \caption{Cascade error (tests run, multiple fail).}
        \label{fig: cascade_errors}
        \end{figure}
}

\title{MultiFileTest: A Multi-File-Level LLM Unit Test Generation Benchmark and Impact of Error Fixing Mechanisms}

\author{
    { \textbf{Yibo Wang}$^1$ \quad \textbf{Congying Xia}$^2$ \quad \textbf{Wenting Zhao}$^3$} \quad \textbf{Jiangshu Du}$^1$\thanks{Work Done Prior to Amazon}\\
    { \textbf{Chunyu Miao}$^1$ \quad \textbf{Zhongfen Deng}$^1$\footnotemark[1] \quad \textbf{Philip S. Yu}$^1$ \quad \textbf{Chen Xing}$^4$} \\
    {
    $^1$University of Illinois Chicago, 
    $^3$ Salesforce Research, 
    $^4$ Scale AI} \\
    {\texttt{\{ywang633, jdu25, cmiao8, zdeng21, psyu\}@uic.edu}}\\
    {\texttt{congyingxia3@gmail.com, wenting.zhao@salesforce.com}, chen.xing@scale.com}
}
\begin{document}
\maketitle
\begin{abstract}
\input{./sections/0_abstract}
\end{abstract}

\section{Introduction}
\input{./sections/1_introduction}

\section{Related Work}
\input{./sections/2_related_work}

\section{Methodology}
\input{./sections/3_methodology}

\section{Experimental Settings}
\input{./sections/4_experimental_settings}

\section{Experiments}

\input{./sections/5_experiments}
\input{./sections/5_error_analyses}

% \section{Futher Analyses}
% \input{sections/6_further_analyses}

\section{Conclusion}
\input{./sections/7_conclusion}

\section*{Limitations}
\input{./sections/8_limitations}

% \section*{Acknowledgments}

% \clearpage
\bibliography{custom}
\clearpage
\appendix

\input{./sections/Appendix}
\end{document}

%% file: sections/0_abstract.tex
Unit test generation has become a promising and important Large Language Model (LLM) use case. However, existing evaluation benchmarks for LLM unit test generation focus on function- or class-level code (single-file) rather than more practical and challenging multi-file-level codebases.
To address such a limitation, we propose MultiFileTest, a multi-file-level benchmark for unit test generation covering Python, Java, and JavaScript. MultiFileTest features 20 moderate-sized and high-quality projects per language. We evaluate eleven frontier LLMs on MultiFileTest, and the results show that most frontier LLMs tested exhibit moderate performance on MultiFileTest, highlighting the difficulty of MultiFileTest.
We also conduct a thorough error analysis, which shows that even advanced LLMs, such as Gemini-3.0-Pro, exhibit basic yet critical errors, including executability and cascade errors. Motivated by this observation, we further evaluate all frontier LLMs under manual error-fixing and self-error-fixing scenarios to assess their potential when equipped with error-fixing mechanisms.
Our code and dataset is available at \href{https://github.com/YiboWANG214/ProjectTest}{MultiFileTest}.

%% file: sections/1_introduction.tex
Unit testing plays an important role in software development, helping identify bugs and ensure code quality. 
Unit tests verify whether individual components of a software program work as expected—for example, checking if add(2, 3) returns 5.
Writing unit tests is time-consuming, usually accounting for approximately 15.8\% of software development time~\cite{daka2014survey}. 
Therefore, automated test case generation, like search-based~\cite{fraser2011evosuite, harman2009theoretical}, constraint-based~\citep{xiao2013characteristic}, and random-based~\citep{pacheco2007feedback} methods, has been proposed to create unit tests. However, these methods often produce less readable tests and are limited to certain types of functions~\cite{grano2018empirical}. Recently, Large Language Models (LLMs) have significantly accelerated unit test generation and improved readability and generalizability with little to no human effort~\cite{siddiq2024using, xie2023chatunitest}. 

Despite the rapid adoption of LLMs for unit testing, the evaluation of LLM unit test generation capabilities appears to be lagging behind. 
Existing benchmarks primarily focus on function, class, or single-file level code~\cite{chen2021evaluating, du2023classeval, wang2025testeval, jain2024testgeneval}, while real-world scenarios typically involve multi-file codebases where functions interact across files with complex dependencies. For instance, a function in file A might import and use classes from files B and C, which themselves depend on other modules. To properly test such codebases, LLMs must understand these cross-file dependencies and correctly set up the test environment, making it significantly more complex than testing function, class, or single-file level code.
The only benchmark that briefly explores multi-file testing, DevBench~\cite{li2024devbench}, is designed for breadth across diverse software development tasks rather than depth in multi-file unit testing, with no systematic focus on cross-file dependencies or comprehensive error analysis.

Therefore, we propose MultiFileTest, a new multi-file-level unit test generation benchmark that offers a larger, higher-quality project set along with comprehensive error analysis of state-of-the-art LLMs.
MultiFileTest covers three programming languages: Python, Java, and JavaScript. For each language, we construct 20 self-contained multi-file projects from GitHub using clear filtering criteria: moderate-sized projects with multiple files and dependencies between them, each under 1,600 lines of code (fitting within input constraints of most code language models), with quality ensured through substantial stars and forks. 
Compared to DevBench's 16 multi-file unit testing projects (6 Python, 5 Java, and 5 C/C++ projects are multi‑file), MultiFileTest provides 3.75× more multi-file targets (60 projects). MultiFileTest fundamentally differs by enforcing cross-file dependencies in every repository—ensuring multi-file reasoning is a guaranteed property rather than an incidental feature.
This carefully constructed benchmark enables comprehensive evaluation of LLMs' capabilities in handling realistic multi-file testing scenarios.

Our evaluation of eleven frontier LLMs (including Gemini-3.0-Pro~\cite{team2024gemini} and GPT-o1) reveals moderate performance across models, highlighting the difficulty of MultiFileTest. 
We observe that different LLMs exhibit different language-level expertise: Gemini-3.0-Pro ranks first in Python and Java, while GPT-o1 ranks first in JavaScript.
Among three programming languages, Java is the most difficult, primarily due to its stricter syntax. Among all tested models, Gemini-3.0-Pro performs best overall. 

Error analysis reveals that even advanced LLMs, such as Gemini-3.0-Pro, produce significant executability and cascade errors (defined in \S\ref{unit_test_generation}). These errors often stem from misunderstandings of contextual dependencies and program structure—areas where reasoning capabilities of LLMs are critical. Although these errors appear to be preliminary and may be relatively easy to fix, they prevent us from observing more advanced aspects of LLM performance on unit test generation, such as correctness and coverage.
To address this, we manually fix LLM's executability and cascade errors and then re-evaluate the fixed unit tests. 
This allows us to measure both raw performance and potential improvement when combined with error-fixing mechanisms. 
By incorporating error-fixing, we uncover critical insights into both the effort required to refine generated tests and the various types of errors that occur in unit tests generated by different LLMs.
We observe that model rankings change significantly after manual fixes, revealing substantial differences in error distributions and improvement potential among LLMs.
Inspired by these findings from manual fixes, we also explore using LLMs for self-fixing their errors in generating multi-file-level unit tests. The results show that while LLMs can correct some errors in their generated unit tests, their self-fixing abilities still lag behind the quality and reliability of human fixes.

Our contributions include: (1) the first multi-file level benchmark for unit test generation with evaluation of eleven frontier LLMs, (2) thorough error analysis through manual fixing of executability and cascade errors to reveal model potential, and (3) the first assessment of LLMs' self-fixing capabilities for unit test generation.

%% file: sections/2_related_work.tex
\subsection{Unit Test Generation}
Traditional unit test generation methods employ search-based~\citep{harman2009theoretical, fraser2011evosuite, lukasczyk2022pynguin}, constraint-based~\citep{xiao2013characteristic}, or random-based~\citep{pacheco2007feedback} strategies to construct test suites that maximize code coverage. 
While achieving reasonable coverage, these tests often have lower readability and less meaningfulness compared to developer-written tests, limiting their adoption in real-world scenarios~\citep{almasi2017industrial, grano2019scented}.

Large Language Models (LLMs) have demonstrated strong code generation capabilities~\cite{feng-etal-2020-codebert, wang-etal-2023-codet5}, inspiring their use in automated unit test generation. Recent approaches in LLM-enhanced unit test generation leverage zero-shot strategies~\citep{siddiq2024using}, iterative querying~\citep{schafer2023empirical}, fine-tuning on specialized datasets~\citep{alagarsamy2025enhancing}, adaptive context selection~\citep{xie2023chatunitest}, dynamic scaling~\citep{ma-etal-2025-dynamic}, and focusing on subtle code differences~\citep{dakhel2024effective, li2023nuances}. These methods are evaluated with various metrics, including compilation success, test correctness, coverage, and bug detection, and demonstrate that LLMs can effectively surpass traditional test generation techniques.

\subsection{LLM Unit Test Generation Benchmark}
Current benchmarks for LLM-based unit test generation mainly focus on function-level~\cite{wang2025testeval, villmow-etal-2021-contest}, class-level~\cite{du2023classeval}, or single-file-level code~\cite{jain2024testgeneval}. Multi-file-level software testing benchmarks, on the other hand, often target tasks other than unit test generation. For instance, R2E-Eval1~\cite{jain2024r2e} is designed for equivalent test harnesses generation, SWT-Bench~\cite{mundler2024swt} focuses on fixing specific bugs rather than entire projects, and DevBench~\cite{li2024devbench} centers on software development tasks. 
While DevBench touches on multi-file-level unit testing, its dataset is limited in quantity and varies in quality, especially for C/C\# and Java, with only five projects each. Half of its projects for unit test generation are difficult to track, and most identifiable projects have fewer than 250 Stars and 50 Forks. Moreover, its broad focus prevents comprehensive evaluation and error analysis of LLM-based multi-file-level unit test generation.
We include a detailed comparison with other benchmarks in Appendix~\ref{sec: comparison}.

%% file: sections/3_methodology.tex
\input{./sections/3_figure_overview}
We introduce MultiFileTest dataset collection and preprocessing (\S\ref{data_collection_preprocessing}), evaluation metrics (\S\ref{sec: evaluation}), and the unit test generation pipeline (\S\ref{unit_test_generation}) for evaluating LLMs on MultiFileTest across three unit test generation scenarios.

\subsection{Benchmark Dataset}
\label{data_collection_preprocessing}

\textbf{Dataset Collection. }
Our dataset comprises carefully selected multi-file-level GitHub repositories in Python, Java, and JavaScript.
We establish selection criteria based on three key factors: 1) appropriate size (2-15 files, <1600 lines of code), 2) inter-file dependencies, and 3) reliable sources. 
The size threshold ensures code fits within standard LLM input windows without truncation, enabling fair comparison across models with different context lengths. This approach isolates our core evaluation target—the model's ability to generate unit tests—rather than testing long-context management or external tooling.
We limit our selection to repositories with publicly available licenses, ensuring legality and openness. 
To maintain quality and reliability,
we prioritize projects with high numbers of stars and forks, signaling community approval and widespread usage.
We also extract smaller, self-contained projects from oversized codebases, carefully adjusting them to function independently without relying on the original larger projects.
After applying these criteria, we construct 20 representative projects per programming language. 
Dataset statistics are summarized in Table~\ref{tab: dataset}, with detailed information on dataset sources and project-specific information in Appendix~\ref{appendix: dataset}.

\noindent\textbf{Pre-processing. }
Dataset pre-processing involves several key steps to ensure the projects are well-structured and suitable for testing. 
First, we verify all selected projects for syntax errors despite their reliable sources.
Second, for projects extracted from larger codebases, we modify them to be self-contained by reorganizing files, adjusting domain naming conventions, and/or modifying import paths to remove dependencies on external modules.
Next, to enhance the accuracy of line coverage measurements, we consolidate statements that span multiple lines into a single line, ensuring more valid metrics. 
Additionally, we maintain original coding styles as much as possible to preserve diversity across projects, allowing us to assess how LLMs perform when faced with various programming styles.

\input{./sections/1_table_data_statistics}

\subsection{Evaluation Metrics}
\label{sec: evaluation}
We focus on three key aspects when evaluating the generated unit tests: executability rate, correctness rate, and coverage rate.
\textit{Executability rate} (ER) measures the percentage of projects in which the generated test suites compile/execute successfully, indicating how often LLMs produce executable unit test suites. 
The executability rate for all projects in $X$ is defined as 
$ER = \frac{|X^{er}|}{|X|}$, 
where $X$ is the project set and $X^{er}\subset X$ denotes the subset of projects whose test suites compile/execute successfully.
\textit{Correctness rate} (CR) calculates the percentage of unit tests that are correct out of all generated unit tests for each project, providing insight into the accuracy of the test generation process. On average, more than 95\% of vanilla-generated unit tests compare expected and actual values, reinforcing the validity of CR as an evaluation metric. Detailed statistics see Appendix~\ref{appendix: assert_statistics}.
The correctness rate for the project $x$ is defined as
$CR_x = \frac{|T_x^{cor}|}{|T_x|}$, 
where $T_x$ is the generated test suite and $T_x^{cor}\subset T_x$ denotes the correct unit test set for the project $x$.
\textit{Coverage rate} analyzes both line and branch coverage to understand how well the generated unit tests explore the code's functionality.
The coverage rate for the project $x$ is defined as
$CR_x = \frac{covered(x)}{total(x)}$,
where $covered(x)$ denotes the number of covered lines/branches in project $x$ and $total(x)$ the total number of lines/branches in project $x\in X$.

These three evaluation metrics are interdependent. If a project's generated test suite contains executability errors, none of its unit tests can execute successfully, resulting in zero correctness and coverage rates for the project. Additionally, errors causing test failures, such as missing Python dependencies, can also impact coverage rates. Therefore, considering these interdependencies, we extend our analysis beyond vanilla unit test evaluation to include manually fixing these errors. This enables a more comprehensive assessment of LLMs' potential to generate high-quality unit tests once such errors are addressed. This assessment is conducted while maintaining the same quantity and diversity of unit tests originally generated by the LLMs.
Furthermore, we extend our analysis to examine the self-fixing capabilities of LLMs.

\input{./sections/3_figure_data_example}

\subsection{Unit Test Generation }
\label{unit_test_generation}
Figure~\ref{fig: overview} shows an overview of the LLM unit test generation process.
Our unit test generation and evaluation aim to ensure fair and thorough assessments under different scenarios:
\begin{itemize}
[noitemsep,topsep=0.2pt,itemsep=1.8pt, leftmargin=12pt]
    \item \textbf{Scenario 1}: Vanilla unit tests extracted from LLMs' outputs.
    \item \textbf{Scenario 2}: Executable unit tests after manually fixing all executability and cascade errors. 
    \item \textbf{Scenario 3}: Unit tests refined by LLMs self-fixing, provided with error messages and human-LLM conversation history.
\end{itemize}

\noindent\textbf{Scenario 1: Vanilla Unit Test Generation. }
We input the entire project and a carefully crafted prompt into LLMs, ensuring the context and requirements are clearly communicated. 
Complete project codes are provided to ensure LLMs have all the necessary context to generate unit tests for the entire project, as shown in Figure~\ref{fig: dataset}. 
To rigorously evaluate LLM capabilities, we craft language-specific prompts addressing the unique challenges of each programming language. A comprehensive assessment is ensured by requiring LLMs to generate unit tests for all project files and providing targeted instructions on executability rate, correctness rate, and coverage metrics. This methodical prompt engineering significantly enhances the quality and relevance of the LLM-generated outputs.
Appendix~\ref{appendix: prompts} lists all experiment prompts, while Appendix~\ref{sec: ablation} contains the prompt ablation analysis.
Vanilla unit tests are extracted directly from the LLM response based on the input project and prompt. 

\noindent\textbf{Scenario 2: Manual Fixing. }
Manually fixing executability and cascade errors is motivated by empirical observations from scenario 1, where even unit tests generated by state-of-the-art LLMs like Claude-3.5-Sonnet contain significant executability errors, making them non-executable. 
These tests also exhibit cascade errors that, while easily fixable, can impact multiple unit tests or the entire test suite (details in \S~\ref{sec: error analyses}).
Although these errors are preliminary and straightforward to resolve, they obstruct a deeper analysis of other critical aspects of LLM performance in unit test generation, particularly correctness and coverage.

\input{./sections/3_figure_error_example}

Therefore, we apply minimal necessary changes to vanilla unit tests, resolving executability and cascade errors while preserving the original test intent. 
Executability errors prevent test execution entirely. For example, Figure 3 shows a ModuleNotFoundError causing pytest to fail before collecting tests, resulting in zero correctness and coverage. 
Cascade errors occur during test execution and cause cascading failures across multiple unit tests 
due to a single root cause. For instance, Figure 4 shows a missing NumPy import invalidating multiple fundamentally correct tests.

\textbf{Manual Fixing Protocol.} To ensure consistency and minimize annotator bias, we target only mechanical errors with objectively verifiable solutions, detailed in Appendix~\ref{app:fixing_protocol}. These fixes require minimal reasoning—the correct solution is determinable from codebase structure (e.g., import paths from file organization). Fixes average 2-6 lines per project (Table~\ref{tab: manual_fix_LOC}), are validated by successful test execution, and preserve all test logic and assertions. Since the protocol applies uniformly across all models, this scenario enables fair comparative assessment of model potential. Manual fixes are performed by CS PhD candidates specializing in software engineering and program analysis.
Additionally, we evaluate unit tests with only executability errors fixed in Appendix~\ref{appendix: alabtion_compilation}.

\noindent\textbf{Scenario 3: LLM Self-fixing. } 
Inspired by our observation from manual fixing that different LLMs exhibit significantly different potentials after manual fixing, we investigate how LLMs perform in self-fixing on our benchmark. 
We explore LLMs' self-fixing abilities by incorporating human-LLM conversation history and error messages as shown in Figure~\ref{fig: self_fix_prompt}. We provide LLMs with the conversation history (including the system prompt, the user prompt for unit test generation requests, and LLM vanilla response), error messages obtained from the testing framework, and the user prompt for error fixing requests. 
When an open-source LLM's input length is limited, we prioritize the information hierarchically: system prompt, LLM's initial response, error messages, error-fixing requests, and unit test generation requests. 
We truncate less critical information as necessary while reserving at least 2,000 tokens for the LLM's self-fixing outputs.
LLM self-fixing scenario helps us understand LLMs' error-fixing ability and their potential to generate better unit tests when incorporating the self-fixing process (see Appendix~\ref{appendix: albation_multiturn} for design rationale). 
Note that during self-fixing, we do not constrain the target error types to just executability or cascade errors.

\input{./sections/3_figure_self_fix_prompt_Python}

%% file: sections/3_figure_overview.tex
\begin{figure}[t]
     \centering
     \includegraphics[width=0.9\linewidth]{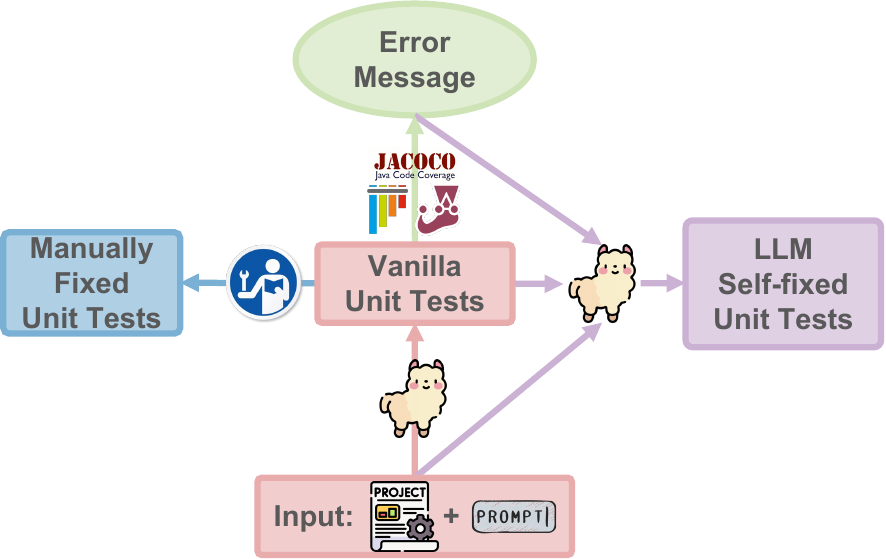}
    \caption{Overview of the unit test generation process.}
    \label{fig: overview}
\end{figure}

%% file: sections/1_table_data_statistics.tex
\begin{table}[t]
\resizebox{\linewidth}{!}{% <-
\begin{tabular}{lccccccc}
\toprule
\textbf{Language} & \textbf{Avg. \#Files} & \textbf{Avg. LOC} & \textbf{Avg. \#Stars} & \textbf{Avg. \#Forks} \\
\midrule
Python &6.10 &654.60	&5810.30	&996.90  \\
Java &4.79	&292.26	&3478.32	&1417.68   \\
JavaScript &4.00	&558.05	&17242.30	&5476.45 \\
\bottomrule
\end{tabular}
}
\caption{Data Statistics (LOC = Lines of Code).}
% \vspace{-15pt}
\label{tab: dataset}
\end{table}

%% file: sections/3_figure_data_example.tex
\begin{dataset}
\end{dataset}

%% file: sections/3_figure_error_example.tex
\begin{compilation_errors}
\end{compilation_errors}
\begin{cascade_errors}
\end{cascade_errors}

%% file: sections/3_figure_self_fix_prompt_Python.tex
\begin{self_fix_prompt}
\scriptsize{
\textbf{System Prompt}: You are a coding assistant... \\
\textbf{User Prompt}: \textit{\{Original Codes\}} Please generate enough unit test cases... \\
\textbf{LLM Response}: \textit{\{Generated Vanilla Unit Tests\}}\\
\textbf{User Prompt}: Here are the error messages from the tests: \textit{\{Error Messages\}}. Errors exist in the generated unit tests. Please fix the unit tests to address these errors and provide the entire unit tests.
}
\end{self_fix_prompt}

%% file: sections/4_experimental_settings.tex
\subsection{Models}
We evaluate seven closed-source models: 
GPT-4-Turbo~\cite{achiam2023gpt}, GPT-3.5-Turbo, GPT-o1, GPT-5-mini, 
Gemini-2.0-Flash-Exp (Gemini-2.0-Flash)~\cite{team2024gemini}, Gemini-3.0-Pro, and
Claude-3.5-Sonnet-20241022 (Claude-3.5-Sonnet)~\cite{anthropic2024claude}, 
and four open-sourced models: CodeQwen1.5-7B-Chat (CodeQwen1.5)~\cite{bai2023qwen}, DeepSeek-Coder-6.7b-Instruct (DeepSeek-Coder)~\cite{guo2024deepseek, zhu2024deepseek}, CodeLlama-7b-Instruct-hf (CodeLlama)~\cite{roziere2023code}, and CodeGemma-7b-it (CodeGemma)~\cite{team2024codegemma}. Detailed information is in Appendix~\ref{appendix: models}.

\subsection{Implementation Details}
We use zero-shot prompting with temperature 0 for unit test generation, running experiments on 8 NVIDIA A100 GPUs with input length maximized to each LLM's token limit.
We use Pytest\footnote{https://docs.pytest.org/en/stable/} for Python, JUnit\footnote{https://junit.org/} for Java, and Jest\footnote{https://jestjs.io/} for JavaScript regarding testing frameworks. For Java code coverage, we use JaCoCo\footnote{https://www.eclemma.org/jacoco/}.

%% file: sections/5_experiments.tex
\input{./sections/5_table_main_results}

We evaluate the generated unit tests from three scenarios, vanilla (\S~\ref{sec: main results}), after manual fixing of executability and cascade errors (\S~\ref{sec: manual fixing}), and LLM self-fixing (\S~\ref{sec: self-fixing}). For each scenario, we evaluate the Correctness Rate (CR), Executability Rate (ER), Line Coverage (LC), and Branch Coverage (BC). We also conduct unique contribution analyses (\S\ref{sec: unique}) and detailed error analyses (\S~\ref{sec: error analyses}).

\subsection{Main Results}
\label{sec: main results}
The main results of the LLMs' unit test generation performance focus on the vanilla unit tests extracted directly from the LLMs' outputs without any changes. This scenario assesses the LLMs' raw capability to generate multi-file-level unit tests.

Table~\ref{tab: main_results} shows the evaluation results for vanilla unit tests.
First, LLMs demonstrate varying language-level expertise. For example, Gemini-3.0-Pro performs the best in Java but falls behind GPT-o1 in JavaScript.

Among three programming languages, Java poses the greatest challenge due to its stricter syntax requirements. Many models fail to generate valid Java code, leading to low executability rates and execution coverage.
Among all the evaluated models, Gemini-3.0-Pro performs the best in general. CodeLlama and CodeGemma have the worst general performance.
We also observe that some models tend to generate more unit tests. However, generating more unit tests does not necessarily lead to better coverage rates. For example, Gemini-2.0-Flash tends to generate the most unit tests but does not obtain the best coverage rate. 
Additionally, sometimes the open-source model can even outperform some closed-source models. For example, DeepSeek-Coder surpasses GPT-3.5-Turbo on Python and JavaScript.
Finally, we confirmed from such results that dependencies exist in metrics. On Java, models like CodeQwen1.5, CodeLlama, and CodeGemma fail to generate executable unit tests, resulting in the lowest correctness and coverage rates. 
We verify the robustness of these experimental results through multiple runs in Appendix~\ref{appendix: robust_analysis}.

\subsection{Manual Fixing Results}
\label{sec: manual fixing}
Table~\ref{tab: manual_results_improvements} presents evaluation results after manual fixing, highlighting substantial improvements compared to vanilla outputs across all programming languages and LLMs. These significant gains demonstrate that LLM-generated unit tests are highly sensitive to executability and cascade errors.

Among programming languages, Java benefits most from manual fixing. In the vanilla scenario, Java exhibits the lowest executability rates, making it particularly challenging. However, after manual fixing, Java shows the most substantial improvement, highlighting the potential of LLMs for Java after fixing executability and cascade errors. 
Among all models, CodeLlama and CodeGemma continue to demonstrate the weakest overall results.
Gemini-2.0-Flash shows the best coverage improvement overall, indicating exceptional potential for better unit test generation once executability and cascade errors are fixed.
Our analysis reveals that manual fixing can reorder model performance rankings. 
For example, in Java, CodeQwen1.5 outperforms DeepSeek-Coder and is now on par with GPT-4-Turbo after fixing. 
In Python, Gemini-2.0-Flash surpasses CodeQwen1.5, showing better potential post-fix. In JavaScript, GPT-3.5-Turbo reaches parity with DeepSeek-Coder.

\input{./sections/5_table_manual_fix}
\subsection{LLMs Self-fixing Results}
\label{sec: self-fixing}
\input{./sections/5_table_self_fix}
LLM self-fixing utilizes human-LLM conversation history and error messages to assist LLMs in fixing errors. This scenario assesses LLMs’ self-fixing capabilities and their potential to generate better unit tests by incorporating self-fixing.

Table~\ref{tab: results_self_fix} shows the LLM self-fixing evaluation results compared with manual fixing results. First, we observe that most closed-source models have effective self-fixing abilities, generating better unit tests than vanilla results. In contrast, the evaluated open-source models lack reliable self-fixing abilities. This limitation likely stems from restricted input length, which leads to incomplete context, alongside weaker comprehension and instruction-following abilities. For instance, CodeGemma and CodeLlama tend to generate textual instructions for fixing errors rather than directly producing the corrected unit tests specified in the prompt.

Second, LLM self-fixing follows similar but not identical trends to manual fixing, suggesting that although LLMs' improvement potential generally aligns with self-fixing capabilities, some LLMs deviate from this pattern. In JavaScript, GPT-o1's self-fixing yields substantially lower coverage rates compared to manual fixing due to generating fewer unit tests and achieving lower executability rates.

Despite currently underperforming compared to manual fixing, LLM self-fixing demonstrates significant potential. Self-fixing has proven effective when LLMs have the necessary capabilities, and it even has the potential to surpass manual fixing due to its flexibility. For example, in JavaScript, CodeQwen1.5 shows greater improvement through self-fixing than manual fixing. 
This occurs because CodeQwen1.5 occasionally misinterprets prompts in vanilla outputs, generating no unit tests.
While manual fixing cannot remedy this fundamental understanding issue, self-fixing enables the model to correctly interpret test generation requirements when error messages indicate missing tests.

\subsection{Unique Contribution of Unit Tests}
\label{sec: unique}
Beyond standard coverage metrics, we introduce unique contribution to assess the efficiency and non-redundancy of generated unit tests. While similar uniqueness metrics have been explored in prior work (e.g., TestPilot~\cite{schafer2023empirical}), we apply this concept to our distinct context of multi-file test generation scenarios.
The unique contribution is defined as the total portion of coverage contributed by each generated unit test that does not overlap with the coverage of other unit tests. 
This measure addresses two critical limitations of conventional metrics. First, it accounts for variations in test quantity across different LLMs, as relying solely on coverage rate becomes insufficient when models produce widely differing numbers of tests. Second, it recognizes the importance of achieving high coverage with minimal tests, as executing numerous tests can be resource-intensive and time-consuming. Further details in Appendix~\ref{sec: more unique}.

Table~\ref{tab: unique_contribution} reveals that all tested LLMs exhibit low unique contribution rates, indicating a tendency toward redundant and repetitive unit tests. 
Although GPT-o1 has better coverage rates than GPT-5-mini, it produces significantly more unit tests, and its unique contribution is lower than GPT-5-mini’s, indicating it prioritizes quantity over quality to attain higher coverage.
GPT-o1 potentially compromises the overall efficiency of the testing process.
\input{./sections/6_table_unique}

%% file: sections/5_table_main_results.tex
\begin{table}[t]
\centering
\resizebox{\linewidth}{!}{
\setlength{\tabcolsep}{3pt}
\begin{tabular}{lcccccc}
\toprule
\textbf{Model} & \textbf{CR} & \textbf{ER} & \textbf{LC} & \textbf{BC} & \textbf{\#Tests} & \textbf{\#Correct} \\
\midrule
\multicolumn{7}{c}{\textbf{Python}} \\
\midrule
GPT-4-Turbo & 47 & 65 & 40 & 36 & 12.60 & 6.15 \\
GPT-3.5-Turbo & 37 & 60 & 38 & 34 & 16.90 & 6.65 \\
GPT-o1 & 60 & 65 & \underline{56} & \underline{54} & 36.35 & 21.7\\
GPT-5-mini & 53 & 60 & 51 & 50 & 12.65 & 7.20 \\
Gemini-2.0-Flash& 46 & 65 & 42 & 39 & 34.95 & 16.95 \\
Gemini-3.0-Pro& \textbf{77} & \textbf{85} & \textbf{76} & \textbf{73} & 27.45 & 21.50 \\
Claude-3.5-Sonnet & \underline{64} & \underline{70} & 51 & 47 & 18.05 & 10.40 \\
CodeQwen1.5 & 24 & 65 & 43 & 40 & 25.40 & 6.80 \\
DeepSeek-Coder & 37 & \underline{70} & 39 & 35 & 7.20 & 2.95 \\
CodeLlama & 16 & 60 & 41 & 37 & 19.30 & 3.95 \\
CodeGemma & 13 & 50 & 31 & 28 & 15.00 & 2.30 \\
\midrule
\multicolumn{7}{c}{\textbf{Java}} \\
\midrule
GPT-4-Turbo & 21 & 35 & 15 & 12 & 7.05 & 2.20 \\
GPT-3.5-Turbo & 13 & 25 & 8 & 7 & 7.50 & 0.80 \\
GPT-o1 & 41 & 60 & 44 & 35 & 15.70 & 6.85 \\
GPT-5-mini & \underline{55} & 70 & \underline{47} & \underline{44} & 11.55 & 7.00 \\
Gemini-2.0-Flash& 19 & 30 & 14 & 12 & 23.30 & 3.90 \\
Gemini-3.0-Pro& \textbf{62} & \textbf{80} & \textbf{58} & \textbf{51} & 13.80 & 10.35 \\
Claude-3.5-Sonnet & 53 & \underline{75} & \underline{47} & 33 & 12.35 & 7.30 \\
CodeQwen1.5 & 0 & 0 & 0 & 0 & 12.95 & 0.00 \\
DeepSeek-Coder & 8 & 20 & 5 & 5 & 7.00 & 0.60 \\
CodeLlama & 0 & 0 & 0 & 0 & 7.85 & 0.00 \\
CodeGemma & 0 & 0 & 0 & 0 & 10.50 & 0.00 \\
\midrule
\multicolumn{7}{c}{\textbf{JavaScript}} \\
\midrule
GPT-4-Turbo & 67 & 75 & 56 & 46 & 16.30 & 11.10 \\
GPT-3.5-Turbo & 51 & 65 & 37 & 28 & 13.25 & 8.05 \\
GPT-o1 & \textbf{87} & \textbf{95} & \textbf{87} & \textbf{75} & 39.40 & 33.30\\
GPT-5-mini & \underline{73} & \underline{90} & \underline{81} & \underline{67} & 18.60 & 14.25 \\
Gemini-2.0-Flash& 59 & 70 & 64 & 61 & 45.85 & 22.55 \\
Gemini-3.0-Pro& 68 & 70 & 68 & 66 & 33.05 & 23.45 \\
Claude-3.5-Sonnet & 65 & 80 & 59 & 53 & 20.25 & 13.35 \\
CodeQwen1.5 & 23 & 35 & 25 & 20 & 8.45 & 4.80 \\
DeepSeek-Coder & 62 & 85 & 50 & 35 & 11.85 & 7.90 \\
CodeLlama & 26 & 85 & 20 & 14 & 48.75 & 18.00 \\
CodeGemma & 29 & 55 & 28 & 21 & 9.00 & 3.00 \\
\bottomrule
\end{tabular}
\setlength{\tabcolsep}{3pt}
}
\caption{Main Results. CR: Correctness Rate (\%), ER: Executability Rate (\%), LC: Line Coverage (\%), BC: Branch Coverage (\%), \#Tests: Avg. \#Tests Per Project.}
\label{tab: main_results}
% \vspace{-10pt}
\end{table}

%% file: sections/5_table_manual_fix.tex
\begin{table}[t]
% \small
\centering
\resizebox{\linewidth}{!}{
\setlength{\tabcolsep}{2.2pt}
\begin{tabular}{lcccccc}
\toprule
\textbf{Model} & \textbf{CR} & \textbf{ER} & \textbf{LC} & \textbf{BC} & \textbf{\#Tests} & \textbf{\#Correct} \\
\midrule
\multicolumn{7}{c}{\textbf{Python}} \\
\midrule
GPT-4-Turbo & 74\textsubscript{(+27)} & 100 & 65\textsubscript{(+25)} & 59\textsubscript{(+23)} & 12.60 & 9.30 \\
GPT-3.5-Turbo & 64\textsubscript{(+27)} & 100 & 63\textsubscript{(+25)} & 57\textsubscript{(+23)} & 16.90 & 10.50 \\
GPT-o1 & 89\textsubscript{(+29)} & 100 & \underline{88}\textsubscript{(+32)} & \underline{86}\textsubscript{(+32)} & 36.35 & 32.25 \\
GPT-5-mini & 83\textsubscript{(+30)} & 100 & 83\textsubscript{(+27)} & 80\textsubscript{(+26)} & 12.65 & 10.80 \\
Gemini-2.0-Flash & 61\textsubscript{(+15)} & 100 & 71\textsubscript{(+29)} & 68\textsubscript{(+29)} & 34.95 & 22.10 \\
Gemini-3.0-Pro& \underline{90}\textsubscript{(+13)} & 100 & \textbf{91}\textsubscript{(+15)} & \textbf{88}\textsubscript{(+15)} & 27.45 & 24.60  \\
Claude-3.5-Sonnet & \textbf{92}\textsubscript{(+28)} & 100 & 74\textsubscript{(+23)} & {70}\textsubscript{(+23)} & 18.05 & 16.40 \\
CodeQwen1.5 & 46\textsubscript{(+22)} & 100 & 70\textsubscript{(+27)} & 65\textsubscript{(+25)} & 25.40 & 10.90 \\
DeepSeek-Coder & 53\textsubscript{(+16)} & 100 & 60\textsubscript{(+21)} & 54\textsubscript{(+19)} & 7.20 & 4.10 \\
CodeLlama & 31\textsubscript{(+15)} & 100 & 61\textsubscript{(+20)} & 56\textsubscript{(+19)} & 19.30 & 7.20 \\
CodeGemma & 36\textsubscript{(+23)} & 100 & 54\textsubscript{(+23)} & 49\textsubscript{(+21)} & 15.00 & 7.85 \\
\midrule
\multicolumn{7}{c}{\textbf{Java}} \\
\midrule
GPT-4-Turbo & 59\textsubscript{(+38)} & 100 & 40\textsubscript{(+25)} & 32\textsubscript{(+20)} & 7.05 & 5.05 \\
GPT-3.5-Turbo & 54\textsubscript{(+41)} & 100 & 36\textsubscript{(+28)} & 27\textsubscript{(+20)} & 7.50 & 4.55 \\
GPT-o1 & {64}\textsubscript{(+23)} & 100 & \textbf{65}\textsubscript{(+21)} & \textbf{56}\textsubscript{(+21)} & 15.7 & 10.75 \\
GPT-5-mini & \underline{71}\textsubscript{(+16)} & 100 & \underline{60}\textsubscript{(+13)} & 52\textsubscript{(+8)} & 11.55 & 8.70 \\
Gemini-2.0-Flash & 56\textsubscript{(+37)} & 100 & 54\textsubscript{(+40)} & {53}\textsubscript{(+41)} & 23.30 & 15.25 \\
Gemini-3.0-Pro& 67\textsubscript{(+5)} & 100 & \underline{60}\textsubscript{(+2)} & \underline{55}\textsubscript{(+4)} & 13.80 & 11.20\\
Claude-3.5-Sonnet & \textbf{74}\textsubscript{(+21)} & 100 & \underline{60}\textsubscript{(+13)} & {53}\textsubscript{(+20)} & 12.35 & 9.65 \\
CodeQwen1.5 & 60\textsubscript{(+60)} & 100 & 42\textsubscript{(+42)} & 31\textsubscript{(+31)} & 12.95 & 8.40 \\
DeepSeek-Coder & 52\textsubscript{(+44)} & 100 & 33\textsubscript{(+28)} & 19\textsubscript{(+14)} & 7.00 & 3.80 \\
CodeLlama & 36\textsubscript{(+36)} & 100 & 25\textsubscript{(+25)} & 20\textsubscript{(+20)} & 7.85 & 4.95 \\
CodeGemma & 57\textsubscript{(+57)} & 100 & 37\textsubscript{(+37)} & 22\textsubscript{(+22)} & 10.50 & 6.50 \\
\midrule
\multicolumn{7}{c}{\textbf{JavaScript}} \\
\midrule
GPT-4-Turbo & {89}\textsubscript{(+22)} & 100 & 75\textsubscript{(+19)} & 59\textsubscript{(+13)} & 16.30 & 14.20 \\
GPT-3.5-Turbo & 74\textsubscript{(+23)} & 100 & 58\textsubscript{(+21)} & 45\textsubscript{(+17)} & 13.25 & 11.20 \\
GPT-o1 & \textbf{91}\textsubscript{(+4)} & 100 & \underline{92}\textsubscript{(+5)} & {79}\textsubscript{(+4)} & 39.40 & 35.15 \\
GPT-5-mini & \underline{90}\textsubscript{(+17)} & 100 & \textbf{94}\textsubscript{(+13)} & \underline{82}\textsubscript{(+15)} & 18.60 & 16.95 \\
Gemini-2.0-Flash & 76\textsubscript{(+17)} & 100 & {88}\textsubscript{(+24)} & {80}\textsubscript{(+19)} & 45.85 & 33.45 \\
Gemini-3.0-Pro& 89\textsubscript{(+21)} & 100 & 88\textsubscript{(+20)} & \textbf{83}\textsubscript{(+17)} & 33.05 & 30.65 \\
Claude-3.5-Sonnet & 87\textsubscript{(+22)} & 100 & 77\textsubscript{(+18)} & 68\textsubscript{(+15)} & 20.25 & 17.55 \\
CodeQwen1.5 & 32\textsubscript{(+9)} & 100 & 35\textsubscript{(+10)} & 27\textsubscript{(+7)} & 8.45 & 6.15 \\
DeepSeek-Coder & 67\textsubscript{(+5)} & 100 & 58\textsubscript{(+8)} & 43\textsubscript{(+8)} & 11.85 & 8.10 \\
CodeLlama & 62\textsubscript{(+36)} & 100 & 44\textsubscript{(+24)} & 28\textsubscript{(+14)} & 48.75 & 31.50 \\
CodeGemma & 58\textsubscript{(+29)} & 100 & 50\textsubscript{(+22)} & 38\textsubscript{(+17)} & 9.00 & 6.40 \\
\bottomrule
\end{tabular}
\setlength{\tabcolsep}{2.2pt}
}
\caption{Manual Fixing Results with Improvements Shown in Parentheses.}
\label{tab: manual_results_improvements}
% \vspace{-8pt}
\end{table}

%% file: sections/5_table_self_fix.tex
\begin{table}[t]
\centering
\resizebox{\linewidth}{!}{
\setlength{\tabcolsep}{2.2pt}
\begin{tabular}{lcccccc}
\toprule
\textbf{Model} & \textbf{CR} & \textbf{ER} & \textbf{LC} & \textbf{BC} & \textbf{\#Tests} & \textbf{\#Correct} \\
\midrule
\multicolumn{7}{c}{\textbf{Python}} \\
\midrule
GPT-4-Turbo & 52\textsubscript{(-22)} & 70\textsubscript{(-30)} & 39\textsubscript{(-26)} & 35\textsubscript{(-24)} & 8.85 & 4.55 \\
GPT-3.5-Turbo & 52\textsubscript{(-12)} & {75}\textsubscript{(-25)} & 45\textsubscript{(-18)} & 39\textsubscript{(-18)} & 14.15 & 8.20 \\
GPT-o1 & {67}\textsubscript{(-22)} & 70\textsubscript{(-30)} & {60}\textsubscript{(-28)} & {58}\textsubscript{(-28)} & 35.50 & 24.35 \\
GPT-5-mini & 66\textsubscript{(-17)} & 70\textsubscript{(-30)} & 61\textsubscript{(-22)} & 59\textsubscript{(-21)} & 12.85 & 9.20 \\
Gemini-2.0-Flash& 47\textsubscript{(-14)} & 60\textsubscript{(-40)} & 45\textsubscript{(-26)} & 42\textsubscript{(-26)} & 34.95 & 17.40 \\
Gemini-3.0-Pro& \textbf{91}\textsubscript{(+1)} & \textbf{95}\textsubscript{(-5)} & \textbf{87}\textsubscript{(-4)} & \textbf{85}\textsubscript{(-3)} & 24.45 & 22.15 \\
Claude-3.5-Sonnet & \underline{86}\textsubscript{(-6)} & \underline{90}\textsubscript{(-10)} & \underline{67}\textsubscript{(-7)} & \underline{63}\textsubscript{(-7)} & 18.00 & 15.55 \\
CodeQwen1.5 & 22\textsubscript{(-24)} & 60\textsubscript{(-40)} & 41\textsubscript{(-29)} & 37\textsubscript{(-28)} & 25.15 & 6.25 \\
DeepSeek-Coder & 18\textsubscript{(-35)} & 35\textsubscript{(-65)} & 20\textsubscript{(-40)} & 18\textsubscript{(-36)} & 4.30 & 1.45 \\
CodeLlama & 0\textsubscript{(-31)} & 5\textsubscript{(-95)} & 5\textsubscript{(-56)} & 5\textsubscript{(-51)} & 3.90 & 0.00 \\
CodeGemma & 8\textsubscript{(-28)} & 25\textsubscript{(-75)} & 14\textsubscript{(-40)} & 13\textsubscript{(-36)} & 9.15 & 0.70 \\
\midrule
\multicolumn{7}{c}{\textbf{Java}} \\
\midrule
GPT-4-Turbo & 43\textsubscript{(-16)} & 55\textsubscript{(-45)} & 26\textsubscript{(-14)} & 18\textsubscript{(-14)} & 6.40 & 2.80 \\
GPT-3.5-Turbo & 17\textsubscript{(-37)} & 25\textsubscript{(-75)} & 11\textsubscript{(-25)} & 12\textsubscript{(-15)} & 6.90 & 1.05 \\
GPT-o1 & \textbf{68}\textsubscript{(+4)} & \textbf{85}\textsubscript{(-15)} & \underline{58}\textsubscript{(-7)} & \underline{54}\textsubscript{(-2)} & 15.60 & 10.10 \\
GPT-5-mini & 61\textsubscript{(-10)} & 70\textsubscript{(-30)} & 47\textsubscript{(-13)} & 45\textsubscript{(-7)} & 9.15 & 6.65 \\
Gemini-2.0-Flash& 31\textsubscript{(-25)} & 40\textsubscript{(-60)} & 29\textsubscript{(-25)} & 24\textsubscript{(-29)} & 22.65 & 7.15 \\
Gemini-3.0-Pro& \underline{67}\textsubscript{(-0)} & \underline{75}\textsubscript{(-25)} & \textbf{61}\textsubscript{(+1)} & \textbf{57}\textsubscript{(+2)} & 13.90 & 11.45 \\
Claude-3.5-Sonnet & {55}\textsubscript{(-19)} & {70}\textsubscript{(-30)} & {39}\textsubscript{(-21)} & {31}\textsubscript{(-22)} & 10.95 & 6.70 \\
CodeQwen1.5 & 5\textsubscript{(-55)} & 5\textsubscript{(-95)} & 0\textsubscript{(-42)} & 0\textsubscript{(-31)} & 12.60 & 0.05 \\
DeepSeek-Coder & 13\textsubscript{(-39)} & 20\textsubscript{(-80)} & 5\textsubscript{(-28)} & 2\textsubscript{(-17)} & 1.35 & 0.25 \\
CodeLlama & 0\textsubscript{(-36)} & 0\textsubscript{(-100)} & 0\textsubscript{(-25)} & 0\textsubscript{(-20)} & 1.30 & 0.00 \\
CodeGemma & 2\textsubscript{(-55)} & 5\textsubscript{(-95)} & 3\textsubscript{(-34)} & 0\textsubscript{(-22)} & 1.75 & 0.05 \\
\midrule
\multicolumn{7}{c}{\textbf{JavaScript}} \\
\midrule
GPT-4-Turbo & 70\textsubscript{(-19)} & {85}\textsubscript{(-15)} & 48\textsubscript{(-27)} & 35\textsubscript{(-24)} & 8.35 & 6.35 \\
GPT-3.5-Turbo & 64\textsubscript{(-10)} & 75\textsubscript{(-25)} & 40\textsubscript{(-18)} & 30\textsubscript{(-15)} & 9.70 & 5.00 \\
GPT-o1 & 54\textsubscript{(-37)} & 65\textsubscript{(-35)} & 47\textsubscript{(-45)} & 38\textsubscript{(-41)} & 20.30 & 12.25 \\
GPT-5-mini & \textbf{85}\textsubscript{(-5)} & \underline{90}\textsubscript{(-10)} & \textbf{85}\textsubscript{(-9)} & \underline{70}\textsubscript{(-12)} & 19.35 & 17.30 \\
Gemini-2.0-Flash& {75}\textsubscript{(-1)} & {85}\textsubscript{(-15)} & {71}\textsubscript{(-17)} & {65}\textsubscript{(-15)} & 40.95 & 28.65 \\
Gemini-3.0-Pro& \underline{79}\textsubscript{(-10)} & 75\textsubscript{(-25)} & \underline{79}\textsubscript{(-9)} & \textbf{75}\textsubscript{(-8)} & 36.30 & 29.25 \\
Claude-3.5-Sonnet & {74}\textsubscript{(-13)} & 80\textsubscript{(-20)} & 60\textsubscript{(-17)} & {53}\textsubscript{(-15)} & 18.05 & 13.35 \\
CodeQwen1.5 & 55\textsubscript{(+23)} & \textbf{95}\textsubscript{(-5)} & {66}\textsubscript{(+31)} & 52\textsubscript{(+25)} & 26.10 & 15.50 \\
DeepSeek-Coder & 14\textsubscript{(-53)} & 35\textsubscript{(-65)} & 15\textsubscript{(-43)} & 10\textsubscript{(-33)} & 2.90 & 1.00 \\
CodeLlama & 9\textsubscript{(-53)} & 35\textsubscript{(-65)} & 7\textsubscript{(-37)} & 5\textsubscript{(-23)} & 7.15 & 0.55 \\
CodeGemma & 31\textsubscript{(-27)} & 60\textsubscript{(-40)} & 29\textsubscript{(-21)} & 21\textsubscript{(-17)} & 10.85 & 3.05 \\
\bottomrule
\end{tabular}
\setlength{\tabcolsep}{2.2pt}
}
\caption{Evaluation Results after Self-fixing. The Comparisons with Manual Fixing are Shown in Parentheses.}
\label{tab: results_self_fix}
% \vspace{-8pt}
\end{table}

%% file: sections/6_table_unique.tex
\begin{table}[t]
% \resizebox{\linewidth}{!}{% <-
\centering
\small
\begin{tabular}{lcccc}
\toprule
\textbf{Model} & \textbf{\#Tests} & \textbf{LC} & \textbf{BC} & \textbf{Unique} \\
\midrule
GPT-4-Turbo & 12.60 & 40 & 36 & 6.35 \\
GPT-3.5-Turbo & 16.90 & 38 & 34 & 5.90 \\
GPT-o1 & 36.35 & 56 & 54 & 6.75\\
GPT-5-mini & 12.65 & 51 & 50 & \underline{15.10} \\
Gemini-2.0-Flash & 34.95 & 42 & 39 & 6.05 \\
Gemini-3.0-Pro & 21.50 & 76 & 73 & \textbf{17.05} \\
Claude-3.5-Sonnet & 18.05 & 51 & 47 & {11.40} \\
CodeQwen1.5 & 25.40 & 43 & 40 & 3.75 \\
DeepSeek-Coder & 7.20 & 39 & 35 & {8.90} \\
CodeLlama & 19.30 & 41 & 37 & 5.55 \\
CodeGemma & 15.00 & 31 & 28 & 2.70 \\
\bottomrule
\end{tabular}
% }
\caption{Unique Contribution on Vanilla Unit Tests.}
\label{tab: unique_contribution}
\end{table}

%% file: sections/5_error_analyses.tex
\subsection{Error Analyses}
\label{sec: error analyses}
We analyze executability, cascade, and post-fix errors per programming language, identifying common errors and their underlying causes. Full analyses in Appendix~\ref{sec: full_error_analyses}.

\noindent\textbf{Executability Error Analyses. }
In \textbf{\textit{Python}}, common executability errors include incorrect import paths for project functions/classes, hallucinated import names/paths, and mismatched parentheses.
\textbf{\textit{Java}}, being more syntax-heavy, faces various executability errors, like hallucinated methods/constructors/classes, missing essential elements like package declarations, illegal access to private/protected elements, invalid code generation, and improper use of mocking frameworks, along with argument type mismatches, ambiguous references, and incompatible types.
\textbf{\textit{JavaScript}} errors typically include hallucinated imports with incorrect paths, empty test suites, and syntax errors from incomplete code generation or mismatched parentheses.

\noindent\textbf{Cascade Error Analyses. }
\textbf{\textit{Python}} cascade errors include missing imports (e.g., numpy, unittest, project functions/classes) and FileNotFoundError from unmocked external files.
\textbf{\textit{Java}}'s primary cascade error is improper/missing mocking of user interactions, causing unusable coverage reports when tests terminate abruptly.
\textbf{\textit{JavaScript}} struggles with missing imports (e.g., chai, three, project functions/classes), confusion between named and default imports, and Jest framework compliance issues.

\noindent\textbf{Post-Fix Error Analyses. }
Across all languages, mismatches between expected and actual values are the most common error. 
\textbf{\textit{Python}} frequently encounters AttributeError from hallucinated attributes.
\textbf{\textit{Java}} suffers from
NullPointerException, zero interactions with mocks, and failures to release mocks due to improper usage.
\textbf{\textit{JavaScript}} commonly faces TypeError, typically caused by LLMs hallucinating non-existent functions and constructors or LLMs invalidly mocking some variables.

\noindent\textbf{Overall. }
Persistent errors across languages include
hallucinations of functions or classes and missing required functions or classes.
Missing required functions or classes often occurs because LLMs \textit{prioritize logical structure over boilerplate code} and \textit{fail to understand the codebase structure and the dependencies between functions, classes, or modules}, which highlights the significant gap between LLM unit test generation at function/class/single-file levels and at multi-file level. Failure to understand the codebase structure and dependencies can further cause issues like confusing non-package and package-based projects (Python) or incorrectly using functions, classes, or packages (Java).
The most common post-fix error is the mismatch between expected and received values, often caused by incorrect expected values due to the \textit{weak reasoning abilities} of LLMs.

%% file: sections/7_conclusion.tex
In conclusion, we build a reliable and high-quality multi-file-level unit test generation benchmark -- MultiFileTest -- with three programming languages. We comprehensively evaluate eleven LLMs' unit test generation abilities with/without manual fixing and LLM self-fixing mechanism on MultiFileTest. Besides, we conduct comprehensive error analyses per programming language.

%% file: sections/8_limitations.tex
Our study has several limitations. 
First, our focus is primarily on three programming languages—Python, Java, and JavaScript—excluding other relevant languages such as C and C\#.

Second, the scale of projects in our benchmark is limited to approximately 1600 lines of code, which is smaller than many production-scale codebases. This constraint stems from the inherent input length restrictions and context window limitations of current LLMs, which make processing very large codebases impractical for tasks like unit test generation without introducing confounding variables.
Despite this size constraint, these projects are designed to retain key structural characteristics of larger codebases, including multiple files with meaningful inter-file dependencies, cross-file function calls, class inheritance, and shared utility components. This ensures the benchmark still evaluates reasoning across files, which is central to multi-file-level unit test generation.
Our experimental results demonstrate that even at this reduced scale, multi-file-level unit test generation remains challenging for state-of-the-art models like Claude-3.5-Sonnet. Expanding to significantly larger codebases would likely shift the evaluation focus toward context handling techniques (e.g., truncation, retrieval, or hierarchical methods) rather than core LLM test generation ability.
While our benchmark does not represent the full complexity of production systems, it serves as a meaningful and challenging step toward that goal, providing valuable evaluation grounded in the practical capabilities of current LLMs.

%% file: sections/Appendix.tex
\section{Dataset}
\label{appendix: dataset}
We provide the detailed information of our datasets in Table~\ref{tab: individual_dataset_py}, Table~\ref{tab: individual_dataset_java}, and Table~\ref{tab: individual_dataset_js}. We provide programming language, project name, license, link, number of stars, number of forks, number of files, and line of codes for each individual project.

The license of "Author Permission" in Table \ref{tab: individual_dataset_java} means that we obtain the usage permission from the author of the corresponding repository\footnote{https://github.com/frandorado/spring-projects/tree/master}.

\section{More Implementation Details}
\subsection{Prompts}
\label{appendix: prompts}
The prompts are displayed in Figure~\ref{fig: prompt}, ~\ref{fig: prompt_java}, ~\ref{fig: prompt_js}, and ~\ref{fig: prompt_comment}.

\subsection{Models}
\label{appendix: models}
The detailed information of models, including license and link, is provided in Table~\ref{tab: models}.

\section{More Experiments and Statistics}
\label{appendix: more_expers_stats}
\subsection{Assert Statistics}
\label{appendix: assert_statistics}
Table~\ref{tab: asser_statistics} presents the percentages of the vanilla-generated unit tests containing comparisons between expected and actual values per language and per model.

\subsection{Robustness Analysis}
\label{appendix: robust_analysis}
To address concerns about statistical robustness, we conduct three independent runs of unit test generation using GPT-3.5-Turbo as shown in Table~\ref{tab:robust_metrics}. The variance across these runs is minimal, indicating that model performance on MultiFileTest is stable and reproducible, further supporting the benchmark's reliability.

We also conduct additional statistical validation using the correctness rate across all nine models and three scenarios as paired observations, as shown in Table~\ref{tab:statistical_comparison}. We find statistically significant differences (all p-values $\leq 0.009$) with large effect sizes (Cohen’s d $> 1.1$) across vanilla, manual, and self-fixing scenarios.

% B1
\input{./sections/3_figure_prompt_python}
\input{./sections/A2_table_prompts}
% B2
\input{./sections/A2_table_models}
% C1
\input{./sections/A3_assert_statistics}
% C2
\input{./sections/A3_robustness_statitics}
% C3
\input{./sections/A5_table_manual_fix_LOC}
% C4
\input{./sections/A3_comparison_other_methods}

\subsection{Changed LOC Statistics of Manual Fixing}
\label{sec: manual_fix_LOC}
We calculated the average number of lines of code (LOC) changed during manual fixing for Python projects across all models in Table~\ref{tab: manual_fix_LOC}.
We observe that the amount of manual edits is modest and consistent across models. These findings suggest that while models frequently produce errors, many are shallow and fixable with minimal human effort, which reinforces the value of human-in-the-loop and LLM-self-fix workflows.

\subsection{Comparison with Other Methods}
\label{appendix: comparison_other_methods}
We use the ChatUnitest~\cite{chen2024chatunitest} Maven Plugin and follow \citet{wang2024hits} to evaluate GPT-3.5-Turbo on Java projects. HITS not only uses iterative debugging, but also uses sophisticated techniques like method slicing to improve unit test performance. These results, as shown in Table~\ref{tab:comparison_other_methods}, further emphasize the difficulty of MultiFileTest, as even the iterative debugging and complex method still achieve low coverage rates. This comparison helps validate our benchmark and encourages further innovation in LLM-driven test generation.

We conduct additional experiments with EvoSuite~\cite{fraser2011evosuite}, a leading search-based test generation tool for Java. Table~\ref{tab:comparison_other_methods} presents the Line Coverage (LC) and Branch Coverage (BC) of EvoSuite compared to GPT-o1 on Java projects. The results show that vanilla LLMs fall behind EvoSuite, while LLM self-fixing has comparable performance with EvoSuite under this multi-file unit test generation setting.

\section{Ablation Study}
\label{appendix: ablation}

\subsection{Ablation Study on Prompts}
\label{sec: ablation}
\input{./sections/6_table_ablation}
We perform a detailed ablation study to analyze the impact of prompts on the performance of unit test generation by LLMs.
As mentioned in \S~\ref{unit_test_generation}, the prompt is composed of programming language-specific requirements (PL), as well as requirements related to the correctness rate (CR), the executability rate (ER), and the coverage rate metrics (Coverage). We ablate each component and analyze the performance of unit test generation of GPT-4-Turbo using different prompts as shown in Table~\ref{tab: ablation}. 
Requirements related to CR and ER can help improve performance in vanilla unit tests. 
Coverage-related requirements are not always beneficial, possibly because a high coverage rate is too abstract for LLMs to interpret effectively.
Programming language-specific requirements improve performance in CR but have the opposite effect on ER, LC, and BC.

Besides, we follow the prompt template from previous work, such as \citet{siddiq2024using}, to move the prompts into comments (e.g., /*...*/). We compare the performance with and without comment signs in Table~\ref{tab: ablation}. Experimental results show that our prompt demonstrates a significant advantage in CR, while the prompt with comment signs exhibits marginal advantages in ER, LC, and BC.

\subsection{Effect of Executability Errors and Cascade Errors}
\label{appendix: alabtion_compilation}
\input{./sections/A4_table_ablation2}
We manually fix only executability errors and evaluate the corrected unit tests in Table~\ref{tab: ablation2}.

By fixing executability errors, Table~\ref{tab: ablation2} shows significant improvements across all programming languages and LLMs compared to Table~\ref{tab: main_results}, indicating that all the programming languages and LLMs are highly sensitive to executability errors.
Comparing Table~\ref{tab: ablation2} with Table~\ref{tab: manual_results_improvements}, we can observe that CodeQwen1.5, CodeGemma, and CodeLlama are more sensitive to cascade errors.
For Java, the changes in Table~\ref{tab: manual_results_improvements} compared to Table~\ref{tab: ablation2} are primarily due to missing or invalid mocks of user interactions\footnote{We consider coverage rates as not applicable when requiring user interactions.} which occur more frequently in unit tests generated by CodeQwen1.5 and CodeGemma. 

\subsection{Ablation Study on Multi-turn Self-fixing}
\label{appendix: albation_multiturn}
\input{./sections/A4_table_ablation3}
To justify our single-turn self-fixing design choice, we conduct an ablation study evaluating multi-turn self-fixing performance. We tested GPT-3.5-Turbo on 10 Python projects across multiple self-fixing iterations.

The results in Table~\ref{tab:multiturn_ablation} show modest improvements over three turns, with signs of coverage rates plateauing by Turn 3. More critically, multiple iterations in multi-file settings introduce confounding variables. Each turn appends conversation history, rapidly exhausting context windows and forcing omission of critical multi-file dependencies. This creates an attribution problem where failures could stem from either the model's inability or context truncation. 

Our single-turn design isolates core error-correction reasoning from context management capabilities. This design principle aligns with our rationale for moderate-sized projects, ensuring that we evaluate the target capability of test generation and error fixing rather than orthogonal factors such as long-context handling and external tooling, which merit separate investigation beyond the scope of this benchmark.

\section{Detailed Error Analyses}
\label{sec: full_error_analyses}
We conduct complex analyses of executability, cascade, and post-fix errors, highlighting the common errors and potential reasons behind the errors.

\paragraph{Executability Error Analyses}
Figure~\ref{fig: errors1} highlights the detailed executability errors that occurred.
One of the most common executability errors in \textbf{\textit{Python}} arises from the LLM's inability to determine whether the project being tested is a package. Specifically, LLMs struggle to recognize the presence or absence of \textit{\_\_init\_\_.py} files, which define a package, leading to confusion between package-based and non-package projects. This inability leads LLM to fail to correctly import functions or classes from the tested project.
Other executability errors include hallucinating the paths or names of imported functions/classes and mismatched parentheses.
\textbf{\textit{Java}}, a syntax-heavy programming language compared to Python and JavaScript, encounters various executability errors, resulting in a significantly lower executability rate than other languages. Java executability errors often arise from issues like hallucinated methods, constructors, or classes, such as incorrect or non-existent imports and references. Missing essential information, such as required functions, classes, or packages, and package declarations, is also a common problem. Errors frequently occur due to illegal access to private or protected elements, invalid code generation (e.g., generating text instead of code), and improper use of mocking frameworks like Mockito, including incorrect objects, missing or misused MockMvc injections, and argument mismatches. Other errors include incorrect usage of other functions, classes, or packages—such as argument type errors, ambiguous references, or incompatible types.
One of the most common executability errors in \textbf{\textit{JavaScript}} is the hallucination of imported functions or classes, where the issue often lies in incorrect paths for the imported functions or classes. CodeQwen1.5 has a particularly common executability error involving invalid generation. This typically occurs due to difficulty understanding the prompt, the need for more specific or detailed code requirements, or the assumption that the code is part of a larger project, leading it to refuse to generate unit tests. Other executability errors include test suites containing empty unit tests and syntax errors caused by incomplete code generation or mismatched parentheses.

\paragraph{Cascade Error Analyses}
Figure~\ref{fig: errors2} highlights the detailed cascade errors that occurred.
For \textbf{\textit{Python}}, the cascade errors include missing imports of commonly used packages such as numpy and unittest, missing imports of functions or classes from the tested project, and FileNotFoundError. 
For \textbf{\textit{Java}}, the most common cascade error is missing or invalid mocking of user interactions. A proper unit test should simulate user interactions through mocking rather than relying on real user inputs. This issue also results in unusable coverage reports for some tested projects, as the error forces an abrupt termination, preventing the generation of coverage data.
For \textbf{\textit{JavaScript}}, the cascade errors include missing imports of commonly used packages such as chai and three, and missing imports of functions or classes from the tested project. Two other common errors specific to JavaScript are that LLMs may confuse named imports with default imports and fail to comply with the Jest framework.

\paragraph{Post-Fix Error Analyses}
Figure~\ref{fig: errors3} highlights the incorrectness reasons after all manual fixes.
For all programming languages, the mismatch between expected and actual values (AssertionError) is the most common error.
Another frequent error in \textbf{\textit{Python}} is AttributeError, typically caused by LLMs hallucinating non-existent attributes.
Other frequent problems in \textbf{\textit{Java}} include NullPointer Errors, zero interactions with mocks, and failures to release mocks, often due to improper mock usage. For projects tested with the Spring framework, errors specific to Spring are also common.
Another frequent error in \textbf{\textit{JavaScript}} is TypeError, mostly caused by LLMs hallucinating non-existent functions and constructors or LLMs invalidly mocking some variables.

\input{./sections/5_figure_error_analyses}

\section{Comparison with Other Benchmarks}
\label{sec: comparison}
\input{./sections/A7_table_comparison}
Table \ref{tab:benchmark_comparison} presents a comprehensive comparison of major code evaluation datasets across multiple dimensions. Among these, MultiFileTest stands out as the first benchmark specifically designed for multi-language, multi-file unit test generation with robust error analysis capabilities. We particularly highlight the distinction between DevBench and MultiFileTest: while DevBench addresses broader software engineering tasks across the entire development lifecycle, MultiFileTest is specifically designed for unit test generation, providing 60 projects (20 per language) compared to DevBench's smaller subset for unit testing. Furthermore, MultiFileTest uniquely offers fine-grained error analysis and both manual fixing and LLM self-fixing mechanisms, which are not present in DevBench. This makes MultiFileTest particularly valuable for evaluating and improving LLMs' capabilities in generating functional test suites for multi-file software projects.

\section{Comparison between Unique Contribution and Other Metrics}
\label{sec: more unique}
While alternative metrics such as test execution time or lines of code provide valuable insights in single-project contexts, they present significant challenges in multi-project benchmarks. The heterogeneous nature of our benchmark—spanning diverse programming languages, project scales, and architectural paradigms—makes these conventional metrics difficult to normalize meaningfully across projects. Test execution times fluctuate based on external dependencies and environmental factors, while code size metrics vary substantially due to languages and coding styles. In contrast, our unique contribution metric offers a project-agnostic measurement framework that maintains consistent interpretability across the entire benchmark suite. It provides a standardized proxy for test utility that transcends project boundaries. This normalized approach enables meaningful cross-project comparisons that would be impractical with traditional metrics, addressing the specific evaluation requirements of diverse multi-project benchmarks.

\section{Discussion on Context Window Limitations}

To address the context window limitation, we identify three primary lines of methods that have emerged in recent research.

The first line focuses on extending context windows to accommodate larger codebases directly. Recent models demonstrate dramatic improvements, expanding from early limits of thousands of tokens to millions by 2024. LongRoPE~\cite{dinglongrope} extends pre-trained LLMs to 2048k tokens with minimal fine-tuning while maintaining performance at shorter context windows. Llama 4 Scout~\cite{meta2025llama} achieves a 10 million token context window. 

The second line of methods employs Retrieval-Augmented Generation (RAG) to provide only important context instead of full context. This approach involves indexing codebase components and dependencies, then dynamically retrieving only the code segments most relevant to the target function for test generation~\cite{lewis2020retrieval, athale2025knowledge, zhang2023repocoder}. This methodology enables scalability while maintaining dependency awareness without overwhelming the context window.

The third approach utilizes hierarchical decomposition to break down large codebases into manageable components~\cite{almorsi2024guided,mundler2024swt}. Often, this is achieved through agent-like methods that employ multi-pass strategies. These agents first analyze the high-level structure, then progressively focus on specific components while maintaining broader context awareness. This allows for the effective handling of larger systems by managing contextual information at different abstraction levels and enabling specialized agents to tackle sub-problems.

While these approaches show promise for scaling to production-size codebases, they introduce confounding variables that would complicate fair evaluation of core LLM test generation capabilities. Our benchmark's constraint to 1600 lines of code enables evaluation without truncation, retrieval strategies, or hierarchical preprocessing, allowing fair comparison across models with different context lengths. This approach isolates our core evaluation target—the model's ability to generate unit tests—rather than testing long-context management or external tooling. The three methods discussed above represent important future directions once foundational test generation capabilities are well-established and benchmarked at the scale our current LLM capabilities can reliably handle.

\input{./sections/A1_table_dataset}

\input{./sections/Manual_Protocol}

%% file: sections/3_figure_prompt_python.tex
\begin{prompt}
\scriptsize{
\textbf{System Prompt}: You are a coding assistant. You generate only source code. \\
\textbf{User Prompt}: \textit{\{Original Codes\}} Please generate enough unit test cases for each Python file in the project. \textcolor{purple}{Ensure that the import path is correct, depending on whether the project is structured as a package.} \textcolor{blue}{Make sure the tests can successfully compile.} \textcolor{orange}{Make sure the tests have correct results.} \textcolor{red}{Try to achieve the highest coverage rate.} 
}
\end{prompt}

%% file: sections/A2_table_prompts.tex
\begin{prompt_java}
\scriptsize{
\textbf{System Prompt}: You are a coding assistant. You generate only source code. \\
\textbf{User Prompt}: \textit{\{Original Codes\}} Please generate enough unit test cases for each java file in the \{method\_signature\} project. Ensure to use mock properly for unit tests. Make sure the tests can successfully compile. Make sure the tests have correct results. Try to achieve the highest coverage rate. }
\end{prompt_java}

\begin{prompt_js}
\scriptsize{
\textbf{System Prompt}: You are a coding assistant. You generate only source code. \\
\textbf{User Prompt}: \textit{\{Original Codes\}} Please generate enough unit test cases for every javascript file in \{method\_signature\} project. Make sure the tests can successfully compile. Make sure the tests have correct results. Try to achieve the highest coverage rate.}
\end{prompt_js}

\begin{prompt_comment}
\scriptsize{
\textbf{System Prompt}: You are a coding assistant. You generate only source code. \\
\textbf{User Prompt}: \textit{\{Original Codes\}} 
\# {classname}\_test.py\textbackslash n 
\# Test class of \{classname\}.\textbackslash n 
\# Please generate enough unit test cases for each python file in the \{method\_signature\} project. Ensure that the import path is correct, depending on whether the project is structured as a package. Make sure the tests can successfully compile. Make sure the tests have correct results. Try to achieve the highest coverage rate. \textbackslash n 
\# class \{classname}\_test\}\textbackslash n 
\end{prompt_comment}

%% file: sections/A2_table_models.tex
%models
%name, license, link
%2 + 4
\begin{table*}[h]
\small
\resizebox{\linewidth}{!}{% <-
\begin{tabular}{llllll}
\toprule
Model Type & Model Name & License & Link \\
\midrule
Close-sourced &GPT-4-Turbo &- &https://platform.openai.com/docs/models/gpt-4\#gpt-4-turbo-and-gpt-4 \\
Close-sourced &GPT-3.5-Turbo &- &https://platform.openai.com/docs/models/gpt-4\#gpt-3-5-turbo \\
Close-sourced & GPT-o1 & - & https://platform.openai.com/docs/models/o1\\
Close-sourced & GPT-5-mini & - & https://platform.openai.com/docs/models/gpt-5-mini\\
Close-sourced & Gemini-2.0-Flash & - & https://ai.google.dev/gemini-api/docs/models/gemini\#gemini-2.0-Flash\\
Close-sourced & Gemini-2.0-Flash & - & https://ai.google.dev/gemini-api/docs/models/gemini\#gemini-3.0-Pro\\
Close-sourced & Claude-3.5-Sonnet & - & https://www.anthropic.com/claude/sonnet \\
\midrule
Open-sourced &CodeQwen1.5-7B-Chat &Tongyi Qianwen LICENSE AGREEMENT &https://huggingface.co/Qwen/CodeQwen1.5-7B-Chat \\
Open-sourced &DeepSeek-Coder-6.7b-Instruct &DEEPSEEK LICENSE AGREEMENT &https://huggingface.co/deepseek-ai/deepseek-coder-6.7b-instruct \\
Open-sourced &CodeLlama-7b-Instruct-hf &LLAMA 2 COMMUNITY LICENSE AGREEMENT	 &https://huggingface.co/codellama/CodeLlama-7b-Instruct-hf \\
Open-sourced &CodeGemma-7b-it &Gemma Terms of Use &https://huggingface.co/google/codegemma-7b-it \\
\bottomrule
\end{tabular}
}
\caption{Model Details.}
\label{tab: models}
\end{table*}

%% file: sections/A3_assert_statistics.tex
\begin{table*}[t]
\resizebox{\linewidth}{!}{
\begin{tabular}{lccccccccccc}
\toprule
\textbf{Model} & \textbf{GPT-4-Turbo} & \textbf{GPT-3.5-Turbo} & \textbf{GPT-o1} & \textbf{GPT-5-mini2} & \textbf{Gemini} & \textbf{Gemini3} & \textbf{Claude3.5} & \textbf{CodeQwen1.5} & \textbf{DeepSeek-Coder} & \textbf{CodeLlama} & \textbf{CodeGemma} \\
\midrule
\textbf{Python} & 98\% & 99\% & 98\% & 96\% & 89\% & 100\% & 99\% & 97\% & 96\% & 99\% & 88\% \\
\textbf{Java} & 97\% & 90\% & 98\% & 95\% & 98\% & 98\% & 97\% & 89\% & 94\% & 85\% & 93\% \\
\textbf{JavaScript} & 100\% & 89\% & 96\% & 85\% & 100\% & 100\% & 100\% & 100\% & 96\% & 86\% & 100\% \\
\bottomrule
\end{tabular}
}
\caption{Percentages of the Vanilla Unit Tests Containing Expected and Actual Value Comparisons.}
\label{tab: asser_statistics}
\end{table*}

%% file: sections/A3_robustness_statitics.tex
\begin{table}[t]
\resizebox{\linewidth}{!}{
\centering
\begin{tabular}{lcccccc}
\toprule
\textbf{Metric} & \textbf{Run 1} & \textbf{Run 2} & \textbf{Run 3} & \textbf{Mean} & \textbf{Variance} & \textbf{Std Dev} \\
\midrule
CR & 0.37 & 0.34 & 0.37 & 0.36 & 0.0003 & 0.0141 \\
ER & 0.60 & 0.65 & 0.65 & 0.633 & 0.0003 & 0.0236 \\
LC & 38\% & 40\% & 39\% & 39\% & 0.0001 & 0.01 \\
BC & 34\% & 37\% & 35\% & 35.3\% & 0.00015 & 0.0122 \\
\bottomrule
\end{tabular}
}
\caption{Performance Metrics across Multiple Runs Using GPT-3.5-Turbo on Python.}
\label{tab:robust_metrics}
\end{table}

\begin{table*}[t]
\resizebox{\linewidth}{!}{%
\centering
\begin{tabular}{llccccc}
\toprule
\textbf{Comparison} & \textbf{Language} & \textbf{Vanilla/Manual Mean} & \textbf{Manual/Self Mean} & \textbf{t-statistic} & \textbf{p-value} & \textbf{Cohen's d} \\
\midrule
Vanilla vs Manual & Python & 38.2\% & 61.8\% & -4.85 & 0.001 & 1.62 \\
Vanilla vs Manual & Java & 17.2\% & 57.9\% & -8.71 & <0.001 & 2.91 \\
Vanilla vs Manual & JavaScript & 52.1\% & 74.0\% & -5.23 & 0.001 & 1.74 \\
Manual vs Self-fix & Python & 61.8\% & 39.1\% & 3.42 & 0.009 & 1.14 \\
Manual vs Self-fix & Java & 57.9\% & 25.9\% & 4.67 & 0.002 & 1.56 \\
Manual vs Self-fix & JavaScript & 74.0\% & 49.6\% & 3.89 & 0.005 & 1.30 \\
\bottomrule
\end{tabular}
}
\caption{Statistical comparison of the Correctness Rate of different scenarios across languages.}
\label{tab:statistical_comparison}
% \vspace{-5pt}
\end{table*}

%% file: sections/A5_table_manual_fix_LOC.tex
\begin{table*}[t]
\centering
\resizebox{\linewidth}{!}{
\begin{tabular}{lccccccccccc}
\toprule
\textbf{Model} & GPT-4 & GPT-3.5 & GPT-o1 & GPT-5-mini & Gemini2 & Gemini3 & Claude3.5 & CodeQwen1.5 & DeepSeek & CodeLlama & CodeGemma \\
\midrule
\textbf{LOC Changed} & 2.45 & 3.35 & 3.15 & 5.85 & 3.15 & 1.80 & 4.05 & 3.40 & 2.35 & 3.0 & 3.40 \\
\bottomrule
\end{tabular}
}
\caption{Lines of code changed during manual fixing for Python projects.}
\label{tab: manual_fix_LOC}
\end{table*}

%% file: sections/A3_comparison_other_methods.tex
\begin{table}[t]
\centering
\resizebox{\linewidth}{!}{
\begin{tabular}{lccccc}
\hline
\textbf{Method} & \textbf{Model} & \textbf{CR} & \textbf{ER} & \textbf{LC} & \textbf{BC} \\
\hline
Vanilla & GPT-3.5-Turbo & 13 & 25 & 8 & 7 \\
Manual fix & GPT-3.5-Turbo & 54 & 100 & 36 & 27 \\
Self-fix & GPT-3.5-Turbo & 17 & 25 & 11 & 12 \\
HITS & GPT-3.5-Turbo & 75 & 80 & 41 & 29 \\
\hline
Vanilla & GPT-o1 & 41 & 60 & 44 & 35 \\
Manual fix & GPT-o1 & 64 & 100 & 65 & 56 \\
Self-fix & GPT-o1 & 68 & 85 & 58 & 54 \\
EvoSuite & - & - & - & 55 & 57 \\
\hline
\end{tabular}
}
\caption{Comparison with Traditional Method EvoSuite, and LLM-based Methods HITS on Java Projects.}
\label{tab:comparison_other_methods}
\end{table}

%% file: sections/6_table_ablation.tex
\begin{table}[t]
\centering
\resizebox{\linewidth}{!}{
\begin{tabular}{llcccccc}
\toprule
\textbf{Phase} & \textbf{Settings} & \textbf{CR} & \textbf{ER} & \textbf{LC} & \textbf{BC} & \textbf{\#Tests} & \textbf{\#Correct} \\
\midrule
\multirow{6}{*}{\textbf{Vanilla}} & Full Prompt & 47 & 65 & 40 & 36 & 12.60 & 6.15 \\
 & w/o CR & 33 $\downarrow$ & 65 & 42 & 38 & 12.75 & 4.75 \\
 & w/o ER & 35 & 63 $\downarrow$ & 41 & 38 & 11.20 & 3.95 \\
 & w/o Coverage & 43 & 75 & 46 $\uparrow$ & 42 $\uparrow$ & 9.80 & 4.20 \\
 & w/o PL & 47 & 75 & 53 & 49 & 9.95 & 4.35 \\
 & w/ Comments & 41 & 65 & 45 & 41 & 10.65 & 4.15 \\
 \midrule
\multirow{6}{*}{\textbf{Manual}} & Full Prompt & 74 & 100 & 65 & 59 & 12.60 & 9.30 \\
 & w/o CR & 76 $\uparrow$ & 100 & 69 & 64 & 12.75 & 9.90 \\
 & w/o ER & 75 & 100 & 70 & 65 & 11.20 & 8.35 \\
 & w/o Coverage & 68 & 100 & 66 $\uparrow$ & 61 $\uparrow$ & 9.80 & 6.75 \\
 & w/o PL & 70 & 100 & 70 & 66 & 9.95 & 6.90 \\
 & w/ Comments & 66 & 100 & 68 & 62 & 10.65 & 7.00 \\
 \bottomrule
\end{tabular}
}
\caption{Ablation Study. The Performance of Unit Test Generation by GPT-4-Turbo Using Different Prompts.}
\label{tab: ablation}
\end{table}

%% file: sections/A4_table_ablation2.tex
\begin{table}[t]
\centering
\resizebox{\linewidth}{!}{
\begin{tabular}{lcccccc}
\toprule
\textbf{Model} & \textbf{CR} & \textbf{ER} & \textbf{LC} & \textbf{BC} & \textbf{\#Tests} & \textbf{\#Correct} \\
\midrule
\multicolumn{7}{c}{\textbf{Python}} \\
\midrule
GPT-4-Turbo & 73 & 100 & 65 & 59 & 12.60 & 9.10 \\
GPT-3.5-Turbo & 63 & 100 & 62 & 56 & 16.90 & 10.40 \\
GPT-o1 & {89} & 100 & \underline{88} & \underline{85} & 36.35 & 32.25 \\
GPT-5-mini & 83 & 100 & 83 & 80 & 12.65 & 10.80 \\
Gemini-2.0-Flash & 61 & 100 & 71 & 68 & 34.95 & 22.10 \\
Gemini-3.0-Pro & \underline{90} & 100 & \textbf{91} & \textbf{88} & 27.45 & 24.60 \\
Claude-3.5-Sonnet & \textbf{92} & 100 & {74} & {70} & 18.05 & 16.40 \\
CodeQwen1.5 & 40 & 100 & 65 & 59 & 25.40 & 9.60 \\
DeepSeek-Coder & 53 & 100 & 60 & 54 & 7.20 & 4.10 \\
CodeLlama & 26 & 100 & 56 & 50 & 19.30 & 6.15 \\
CodeGemma & 30 & 100 & 52 & 47 & 15.00 & 6.15 \\
\midrule
\multicolumn{7}{c}{\textbf{Java}} \\
\midrule
GPT-4-Turbo & 59 & 100 & 42 & 34 & 7.05 & 5.05 \\
GPT-3.5-Turbo & 48 & 100 & 37 & 29 & 7.50 & 4.20 \\
GPT-o1 & {62} & 100 & \textbf{67} & \underline{56} & 15.70 & 10.50 \\
GPT-5-mini & \underline{71} & 100 & 60 & 52 & 11.55 & 8.70 \\
Gemini-2.0-Flash & 55 & 100 & 54 & 53 & 23.30 & 15.00 \\
Gemini-3.0-Pro & 67 & 100 & 60 & 55 & 13.80 & 11.20 \\
Claude-3.5-Sonnet & \textbf{73} & 100 & \underline{63} & \textbf{57} & 12.35 & 9.60 \\
CodeQwen1.5 & 49 & 100 & 49 & 39 & 12.95 & 7.50 \\
DeepSeek-Coder & 40 & 100 & 36 & 19 & 7.00 & 2.85 \\
CodeLlama & 30 & 100 & 26 & 21 & 7.85 & 4.25 \\
CodeGemma & 46 & 100 & 44 & 26 & 10.50 & 5.55 \\
\midrule
\multicolumn{7}{c}{\textbf{JavaScript}} \\
\midrule
GPT-4-Turbo & {89} & 100 & 75 & 59 & 16.30 & 14.15 \\
GPT-3.5-Turbo & 71 & 100 & 56 & 44 & 13.25 & 10.65 \\
GPT-o1 & \textbf{91} & 100 & \underline{92} & {79} & 39.40 & 35.15 \\
GPT-5-mini & \underline{90} & 100 & \textbf{94} & \textbf{82} & 18.60 & 16.95 \\
Gemini-2.0-Flash & 76 & 100 & {88} & \underline{80} & 45.85 & 33.30 \\
Gemini-3.0-Pro & 89 & 100 & 88 & 83 & 33.05 & 30.65 \\
Claude-3.5-Sonnet & 83 & 100 & 75 & 66 & 20.25 & 16.75 \\
CodeQwen1.5 & 28 & 100 & 29 & 22 & 8.45 & 5.65 \\
DeepSeek-Coder & 66 & 100 & 58 & 43 & 11.85 & 8.05 \\
CodeLlama & 28 & 100 & 20 & 15 & 48.75 & 21.40 \\
CodeGemma & 45 & 100 & 43 & 30 & 9.00 & 5.75 \\
\bottomrule
\end{tabular}
}
\caption{Evaluation Results When Only Manually Fixing Executability Errors.}
\label{tab: ablation2}
\end{table}

%% file: sections/A4_table_ablation3.tex
\begin{table}[h]
\centering
\begin{tabular}{lcccc}
\toprule
Turn & CR & ER & LC & BC \\
\midrule
Turn 1 & 31 & 60 & 36 & 33 \\
Turn 2 & 32 & 70 & 45 & 42 \\
Turn 3 & 34 & 80 & 49 & 45 \\
\bottomrule
\end{tabular}
\caption{Multi-turn Self-fixing Performance across Iterations on GPT-3.5-Turbo on 10 Python Projects}
\label{tab:multiturn_ablation}
\end{table}

%% file: sections/5_figure_error_analyses.tex
\tikzstyle{my-box}=[
    rectangle,
    draw=hidden-draw,
    rounded corners,
    text opacity=1,
    minimum height=1.5em,
    minimum width=5em,
    inner sep=2pt,
    align=center,
    fill opacity=.5,
    line width=0.8pt,
]
\tikzstyle{leaf}=[my-box, minimum height=1.5em,
    fill=hidden-pink!80, text=black, align=left,font=\normalsize,
    inner xsep=2pt,
    inner ysep=4pt,
    line width=0.8pt,
]
\begin{figure}[t]
    \centering
    \resizebox{\linewidth}{!}{
        \begin{forest}
            forked edges,
            for tree={
                grow'=0,
                draw,
                reversed=true,
                anchor=base west,
                parent anchor=east,
                child anchor=west,
                base=left,
                font=\large,
                rectangle,
                rounded corners,
                align=left,
                minimum width=4em,
                edge+={darkgray, line width=1pt},
                s sep=3pt,
                inner xsep=2pt,
                inner ysep=3pt,
                line width=0.8pt,
                ver/.style={rotate=90, child anchor=north, parent anchor=south, anchor=center},
            },
            where level=1{text width=4.4em,font=\normalsize,}{},
            where level=2{text width=12em,font=\normalsize,}{},
            where level=3{text width=25em,font=\normalsize,}{},
            % where level=4{text width=5em,font=\normalsize,}{},
			[
			    Executability Error Analysis, ver
			    [
		              Python, 
                        fill=lgreen
    			            [
                                Confuse between non-package and package-based projects
                                , leaf, text width=28em, fill=lgreen
    			            ]
                                [
                                Hallucinate the imported functions/classes:\\
                                1. Paths of the imported functions/classes are wrong\\
                                2. Names of the imported functions/classes are wrong
                                , leaf, text width=28em, fill=lgreen
    			            ]
                                [
                                Syntax Error: Mismatched parentheses
                                , leaf, text width=28em, fill=lgreen
                                ]
			        ]
			    [
    			      Java, fill=lblue
    			            [
                                Hallucinate methods/constructors/functions/classes:\\
                                1. Paths of the imported functions/classes are wrong \\
                                2. Names of the imported functions/classes are wrong \\
                                3. Non-existed methods/constructors
                                , leaf, text width=28em, fill=lblue
    			            ]
                                [
                                Missing information: \\
                                1. Required functions/classes/packages are missing \\
                                2. Required package information is missing \\
                                3. Unreported exception \\
                                , leaf, text width=28em, fill=lblue
                                ]
                                [
                                Illegal access to private/protected functions/classes
                                , leaf, text width=28em, fill=lblue
                                ]
                                [
                                Invalid generation: \\
                                1. Generate textual instructions instead of codes
                                2. Block by model \\
                                , leaf, text width=28em, fill=lblue
                                ]
                                [
                                Incorrect use of mocking: \\
                                1. Wrong objects provided to Mockito \\
                                2. Missing MockMvc injection 
                                3. Inappropriate mockmvc \\
                                4. Argument mismatch
                                , leaf, text width=28em, fill=lblue
                                ]
                                [
                                Incorrect use of other functions/classes/packages: \\
                                1. Arguments type error 2. Ambiguous reference \\
                                3. Incompatible types
                                , leaf, text width=28em, fill=lblue
                                ]
			    ]
                    [
                        JavaScript, fill=lyellow[
                                Hallucinate the imported functions/classes: \\
                                1. Paths of the imported functions/classes are wrong
                                , leaf, text width=28em, fill=lyellow
                                ]
                                [
                                Invalid generation: \\
                                1. Cannot understand the prompt
                                2. Require more/specific codes \\
                                3. Assume the codes are part of a larger project and \\ decline to generate unit tests
                                , leaf, text width=28em, fill=lyellow
                                ]
                                [
                                Test suits have empty unit tests \\
                                , leaf, text width=28em, fill=lyellow
                                ]
                                [
                                Syntax Error: \\
                                1. Incomplete generation 
                                2. Mismatched parentheses
                                , leaf, text width=28em, fill=lyellow
                                ]
                    ]
			]
            \end{forest}
    }
    \caption{Frequent Executability Errors in Main Results.}
    \label{fig: errors1}
\end{figure}

\tikzstyle{my-box}=[
    rectangle,
    draw=hidden-draw,
    rounded corners,
    text opacity=1,
    minimum height=1.5em,
    minimum width=5em,
    inner sep=2pt,
    align=center,
    fill opacity=.5,
    line width=0.8pt,
]
\tikzstyle{leaf}=[my-box, minimum height=1.5em,
    fill=hidden-pink!80, text=black, align=left,font=\normalsize,
    inner xsep=2pt,
    inner ysep=4pt,
    line width=0.8pt,
]
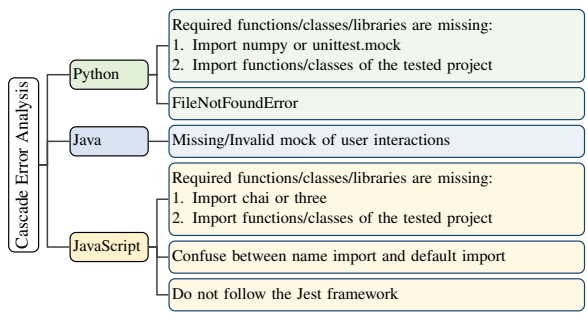
\begin{figure}[t]
    \centering
    \resizebox{\linewidth}{!}{
        \begin{forest}
            forked edges,
            for tree={
                grow'=0,
                draw,
                reversed=true,
                anchor=base west,
                parent anchor=east,
                child anchor=west,
                base=left,
                font=\large,
                rectangle,
                rounded corners,
                align=left,
                minimum width=4em,
                edge+={darkgray, line width=1pt},
                s sep=3pt,
                inner xsep=2pt,
                inner ysep=3pt,
                line width=0.8pt,
                ver/.style={rotate=90, child anchor=north, parent anchor=south, anchor=center},
            },
            where level=1{text width=4.4em,font=\normalsize,}{},
            where level=2{text width=12em,font=\normalsize,}{},
            where level=3{text width=20em,font=\normalsize,}{},
            % where level=4{text width=5em,font=\normalsize,}{},
			[
			    Cascade Error Analysis, ver
			    [
		              Python, 
                        fill=lgreen
    			            [
                                Required functions/classes/libraries are missing:\\
                                1. Import numpy or unittest.mock\\
                                2. Import functions/classes of the tested project
                                , leaf, text width=25em, fill=lgreen
    			            ]
    			            [
                                FileNotFoundError
                                , leaf, text width=25em, fill=lgreen
    			            ]
			        ]
			    [
    			      Java, fill=lblue
    			            [
                                Missing/Invalid mock of user interactions
                                , leaf, text width=25em, fill=lblue
    			            ]
			    ]
                    [
                        JavaScript, fill=lyellow
                                [
                                Required functions/classes/libraries are missing:\\
                                1. Import chai or three\\
                                2. Import functions/classes of the tested project
                                , leaf, text width=25em, fill=lyellow
                                ]
                                [
                                Confuse between name import and default import
                                , leaf, text width=25em, fill=lyellow
                                ]
                                [
                                Do not follow the Jest framework
                                , leaf, text width=25em, fill=lyellow
                                ]
                    ]
			]
            \end{forest}
    }
    \caption{Frequent Cascade Errors.}
    \label{fig: errors2}
\end{figure}

\tikzstyle{my-box}=[
    rectangle,
    draw=hidden-draw,
    rounded corners,
    text opacity=1,
    minimum height=1.5em,
    minimum width=5em,
    inner sep=2pt,
    align=center,
    fill opacity=.5,
    line width=0.8pt,
]
\tikzstyle{leaf}=[my-box, minimum height=1.5em,
    fill=hidden-pink!80, text=black, align=left,font=\normalsize,
    inner xsep=2pt,
    inner ysep=4pt,
    line width=0.8pt,
]
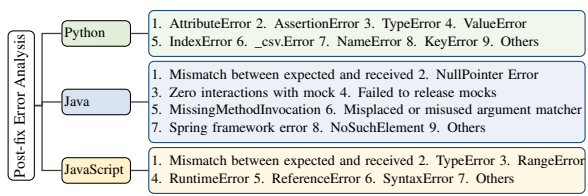
\begin{figure}[t]
    \centering
    \resizebox{\linewidth}{!}{
        \begin{forest}
            forked edges,
            for tree={
                grow'=0,
                draw,
                reversed=true,
                anchor=base west,
                parent anchor=east,
                child anchor=west,
                base=left,
                font=\large,
                rectangle,
                rounded corners,
                align=left,
                minimum width=4em,
                edge+={darkgray, line width=1pt},
                s sep=3pt,
                inner xsep=2pt,
                inner ysep=3pt,
                line width=0.8pt,
                ver/.style={rotate=90, child anchor=north, parent anchor=south, anchor=center},
            },
            where level=1{text width=4.4em,font=\normalsize,}{},
            where level=2{text width=12em,font=\normalsize,}{},
            where level=3{text width=25em,font=\normalsize,}{},
            % where level=4{text width=5em,font=\normalsize,}{},
			[
			    Post-fix Error Analysis, ver
			    [
		              Python, 
                        fill=lgreen
                                [
                                1. AttributeError
                                2. AssertionError
                                3. TypeError 
                                4. ValueError \\
                                5. IndexError 
                                6. \_csv.Error 
                                7. NameError
                                8. KeyError 
                                9. Others
                                , leaf, text width=30em, fill=lgreen
    			            ]
			        ]
			    [
    			      Java, fill=lblue
    			            [
                                1. Mismatch between expected and received 
                                2. NullPointer Error \\
                                3. Zero interactions with mock 
                                4. Failed to release mocks \\
                                5. MissingMethodInvocation 
                                6. Misplaced or misused argument matcher \\
                                7. Spring framework error 
                                8. NoSuchElement 
                                9. Others
                                , leaf, text width=30em, fill=lblue
    			            ]
			    ]
                    [
                        JavaScript, fill=lyellow
                                [
                                1. Mismatch between expected and received 
                                2. TypeError 
                                3. RangeError \\
                                4. RuntimeError
                                5. ReferenceError 
                                6. SyntaxError 
                                7. Others
                                % 7. Invalid component type 
                                % 8. Image given has not completed loading 
                                % 9. Invalid Chai property
                                , leaf, text width=30em, fill=lyellow
                                ]
                    ]
			]
            \end{forest}
    }
    \caption{Frequent Post-Fix Errors.}
    \label{fig: errors3}
\end{figure}

%% file: sections/A7_table_comparison.tex
\begin{table*}[t]
\centering
\resizebox{\linewidth}{!}{
\begin{tabular}{lccccccccc}
\toprule
\textbf{Dataset} & \textbf{Language} & \textbf{Code Level} & \textbf{Multi-file} & \textbf{TestGen} & \textbf{Size} & \textbf{Avg. \#Files} & \textbf{Self-contained} & \textbf{Error Analyses} & \textbf{Error Fixing} \\
\midrule
HumanEval~\cite{chen2021evaluating} & Python & Function & \XSolidBrush & \XSolidBrush & 164 & 1 & \Checkmark & \XSolidBrush & \XSolidBrush \\
ClassEval~\cite{du2023classeval} & Python & Class & \XSolidBrush & \XSolidBrush & 100 & 1 & \Checkmark & \XSolidBrush & \XSolidBrush \\
SWE-bench~\cite{jimenezswe} & Python & Multi-file & \Checkmark & \XSolidBrush & 12 & - & \Checkmark & \XSolidBrush & \XSolidBrush \\
TestEval~\cite{wang2025testeval} & Python & Function & \XSolidBrush & \Checkmark & 210 & 1 & \Checkmark & \XSolidBrush & \XSolidBrush \\
TestGenEval~\cite{jain2024testgeneval} & Python & Single-file & \XSolidBrush & \Checkmark & 1,210 & 1 & \XSolidBrush & \Checkmark & \XSolidBrush \\
DevBench~\cite{li2024devbench} & Python, Java, C/C\# & Multi-file & \Checkmark & \Checkmark\kern-1.1ex\raisebox{1.5ex}{\rotatebox[origin=c]{125}{--}} & 20 & 4.20 & \Checkmark & \XSolidBrush & \XSolidBrush \\
MultiFileTest (ours) & Python, Java, JavaScript & Multi-file & \Checkmark & \Checkmark & 60 & 4.92 & \Checkmark & \Checkmark & \Checkmark \\
\bottomrule
\end{tabular}
}
\caption{Benchmarks comparison. ``TestGen'' refers to whether the benchmark is designed for unit test generation. ``Self-contained'' refers to whether the data sample is independent rather than being part of a larger project. \Checkmark\kern-1.1ex\raisebox{1.5ex}{\rotatebox[origin=c]{125}{--}} indicates partial satisfaction of the condition. ``Error Analyses'' refers to specific error analyses for unit test generation by LLMs.}
\label{tab:benchmark_comparison}
\end{table*}

%% file: sections/A1_table_dataset.tex
\begin{table*}[h]
\small
% \resizebox{\linewidth}{!}{% <-
\centering
\begin{tabular}{llllllllll}
\toprule
 Project Name & License & Link & \#Stars & \#Forks & \#Files & LOC \\
\midrule
blackjack &MIT license &\href{https://github.com/datamllab/rlcard/tree/master/rlcard/games/blackjack}{blackjack} &2937	&641&6&401 \\
bridge &MIT license &\href{https://github.com/datamllab/rlcard/tree/master/rlcard/games/bridge}{bridge} &2937 &641&11&792 \\
doudizhu &MIT license &\href{https://github.com/datamllab/rlcard/tree/master/rlcard/games/doudizhu}{doudizhu} &2937 &641&8&1178 \\
fuzzywuzzy &MIT license &\href{https://github.com/seatgeek/fuzzywuzzy/tree/master/fuzzywuzzy}{fuzzywuzzy} &9200 &876&6&808 \\
gin\_rummy &GPL-2.0 license &\href{https://github.com/datamllab/rlcard/tree/master/rlcard/games/gin_rummy}{gin\_rummy} &2937 &641&14&1453 \\
keras\_preprocessing &MIT license &\href{https://github.com/keras-team/keras-preprocessing/tree/master/keras_preprocessing}{keras\_preprocessing} &1024 &443&3&1002 \\
leducholde &MIT license &\href{https://github.com/datamllab/rlcard/tree/master/rlcard/games/leducholde}{leducholde} &2937 &641&8&833 \\
limitholdem &MIT license &\href{https://github.com/datamllab/rlcard/tree/master/rlcard/games/limitholdem}{limitholdem} &2937 &641&8&1243 \\
mahjong &MIT license &\href{https://github.com/datamllab/rlcard/tree/master/rlcard/games/mahjong}{mahjong} &2937 &641&8&703 \\
nolimitholdem &MIT license &\href{https://github.com/datamllab/rlcard/tree/master/rlcard/games/nolimitholdem}{nolimitholdem} &2937 &641&8&1562 \\
slugify &MIT license &\href{https://github.com/un33k/python-slugify/tree/master/slugify}{slugify} &1500 &109&3&246 \\
stock &CC-BY-SA-4.0 license &\href{https://github.com/dabeaz-course/python-mastery/tree/main/Solutions/7_3}{stock} &10700 &1800&3&217 \\
stock2 &CC-BY-SA-4.0 license &\href{https://github.com/dabeaz-course/python-mastery/tree/main/Solutions/7_6}{stock2} &10700 &1800&5&361 \\
stock3 &CC-BY-SA-4.0 license &\href{https://github.com/dabeaz-course/python-mastery/tree/main/Solutions/8_1}{stock3} &10700 &1800&4&286 \\
stock4 &CC-BY-SA-4.0 license &\href{https://github.com/dabeaz-course/python-mastery/tree/main/Solutions/8_2}{stock4} &10700 &1800&4&323 \\
structly &CC-BY-SA-4.0 license &\href{https://github.com/dabeaz-course/python-mastery/tree/main/Solutions/9_2}{structly} &10700 &1800&6&369 \\
svm &MIT license &\href{https://github.com/rushter/MLAlgorithms/tree/master/mla/svm}{svm} &10800 &1800&4&238 \\
the fuzz &CC-BY-SA-4.0 license &\href{https://github.com/seatgeek/thefuzz/tree/master/thefuzz}{the fuzz} &2949 &141&3&183 \\
tree &CC-BY-SA-4.0 license &\href{https://github.com/rushter/MLAlgorithms/blob/master/mla/ensemble/tree.py}{tree} &10800 &1800&2&225 \\
uno &MIT license &\href{https://github.com/datamllab/rlcard/tree/master/rlcard/games/uno}{uno} &2937 &641&8&669 \\
\bottomrule
\end{tabular}
% }
\caption{Dataset Details (Python).}
\label{tab: individual_dataset_py}
\end{table*}

\begin{table*}[h]
\small
% \resizebox{\linewidth}{!}{% <-
\centering
\begin{tabular}{llllllllll}
\toprule
Project Name & License & Link & \#Stars & \#Forks & \#Files & LOC \\
\midrule
Actor\_relationship\_game & Apache-2.0 license &\href{https://github.com/open-compass/DevEval/tree/main/benchmark_data/java/Actor_relationship_game/src/main/java/Actor_relationship_game}{Actor\_relationship\_game} &85	&5 &7&479\\
banking application &MIT license &\href{https://github.com/kishanrajput23/Java-Projects-Collections/tree/main/banking\%20application}{banking application} &341 &366 &3&357\\
Calculator\-OOPS &MIT license &\href{https://github.com/kishanrajput23/Java-Projects-Collections/tree/main/Calculator-OOPS}{Calculator\-OOPS} &525 &513 &8&138\\
% Email\-Administration\-Application & -&\href{https://github.com/KrishGaur1354/Java-Projects-for-Beginners/tree/main/Email-Administration-Application}{Email\-Administration\-Application} &33 &17 \\
emailgenerator &MIT license &\href{https://github.com/kishanrajput23/Java-Projects-Collections/tree/main/Email_Generator/src/emailgenerator}{emailgenerator} &525 &513 &2&194\\
heap &MIT license &\href{https://github.com/TheAlgorithms/Java/tree/5ab6356090c17cddd953c801eac4abb6ef48c9f1/src/main/java/com/thealgorithms/datastructures/heaps}{heap} &60500 &19600 &3&629\\
idcenter &Apache-2.0 license &\href{https://github.com/adyliu/idcenter}{idcenter} &146 &136 &4&295\\
libraryApp &MIT license &\href{https://github.com/kishanrajput23/Java-Projects-Collections/tree/main/LibraryApp/libraryApp}{libraryApp} &341 &366 &4&340\\
libraryManagement &MIT license &\href{https://github.com/kishanrajput23/Java-Projects-Collections/tree/main/LibraryMangement/src}{libraryManagement} &341 &366 &7&483 \\
logrequestresponseundertow &Author Permission  &\href{https://github.com/frandorado/spring-projects/tree/master/log-request-response-undertow}{logrequestresponseundertow} &152 &131 &3&85\\
Password\_Generator &MIT license &\href{https://github.com/kishanrajput23/Java-Projects-Collections/tree/main/Password_Generator/Password\%20Generator/src}{Password\_Generator} &341 &366 &5&344\\
Pong Game &MIT license &\href{https://github.com/kishanrajput23/Java-Projects-Collections/tree/main/Pong\%20Game}{Pong Game} &341 &366 &6&321\\
redis &Apache-2.0 license &\href{https://github.com/mybatis/redis-cache}{redis} &413 &218 &9&651\\
servlet &MIT license &\href{https://github.com/kishanrajput23/Java-Projects-Collections/tree/main/Online\%20Voting\%20System/Online_Voting_System/src/main/java/vote/com/servlet}{servlet} &341 &366 &6&308\\
simpleChat &MIT license &\href{https://github.com/abhpd/hacktoberfest2021/tree/main/Java/Projects/SimpleChat}{simpleChat} &543 &1500 &2&170\\
springdatamongowithcluster &Author Permission &\href{https://github.com/frandorado/spring-projects/tree/master/spring-data-mongo-with-cluster}{springdatamongowithcluster} &152 &131 &2&51\\
springmicrometerundertow &Author Permission &\href{https://github.com/frandorado/spring-projects/tree/master/spring-micrometer-undertow}{springmicrometerundertow} &152 &131 &3&130\\
springreactivenonreactive &Author Permission &\href{https://github.com/frandorado/spring-projects/tree/master/spring-reactive-nonreactive}{springreactivenonreactive} &152 &131 &7&170\\
springuploads3 &Author Permission &\href{https://github.com/frandorado/spring-projects/tree/master/spring-upload-s3-localstack}{springuploads3} &152 &131 &7&192\\
Train &MIT license &\href{https://github.com/abhpd/hacktoberfest2021/tree/main/Java/Projects/Train}{Train} &545 &1600 &3&216\\
\bottomrule
\end{tabular}
% }
\caption{Dataset Details (Java).}
\label{tab: individual_dataset_java}
\end{table*}

\begin{table*}[h]
\small
% \resizebox{\linewidth}{!}{% <-
\centering
\begin{tabular}{llllllllll}
\toprule
 Project Name & License & Link & \#Stars & \#Forks & \#Files & LOC \\
\midrule
aggregate &MIT license &\href{https://github.com/ehmicky/modern-errors/blob/main/src/merge/aggregate.js}{aggregate} &1500	&18 &2&21\\
animation &MIT license &\href{https://github.com/mrdoob/three.js/blob/dev/src/animation/AnimationAction.js}{animation} &103000 &35400 &2&911\\
check &MIT license &\href{https://github.com/ehmicky/modern-errors/blob/main/src/subclass/check.js}{check} &1500 &18 &2&78\\
circle &MIT license &\href{https://github.com/schteppe/p2.js/blob/master/src/shapes/Circle.js}{circle} &2700 &330 &4&1068\\
ckmeans &ISC license &\href{https://github.com/simple-statistics/simple-statistics/blob/main/src/ckmeans.js}{ckmeans} &3400 &226 &4&364\\
controls &MIT license &\href{https://github.com/mrdoob/three.js/blob/dev/src/extras/Controls.js}{controls} &103000 &35400 &2&119\\
convex &MIT license &\href{https://github.com/schteppe/p2.js/blob/master/src/shapes/Convex.js}{convex} &2700 &330 &6&1878\\
easing &MIT license &\href{https://github.com/alienkitty/space.js/blob/main/src/tween/Easing.js}{easing} &418 &9 &2&341\\
magnetic &MIT license &\href{https://github.com/alienkitty/space.js/blob/main/src/extras/Magnetic.js}{magnetic} &418 &9 &7&949\\
overlapkeeper &MIT license &\href{https://github.com/schteppe/p2.js/blob/master/src/utils/OverlapKeeper.js}{overlapkeeper} &2700 &330 &6&539\\
particle &MIT license &\href{https://github.com/schteppe/p2.js/blob/master/src/shapes/Particle.js}{particle} &2700 &330 &4&963\\
pixelrender &MIT license &\href{https://github.com/drawcall/Proton/blob/master/src/render/PixelRenderer.js}{pixelrender} &2400 &274 &9&794\\
plane &MIT license &\href{https://github.com/schteppe/p2.js/blob/master/src/shapes/Plane.js}{plane} &2700 &330 &4&1052\\
solver &MIT license &\href{https://github.com/schteppe/p2.js/blob/master/src/solver/Solver.js}{solver} &2700 &330 &2&242\\
span &MIT license &\href{https://github.com/drawcall/Proton/blob/master/src/math/Span.js}{span} &2400 &274 &6&536\\
spherical &MIT license &\href{https://github.com/mrdoob/three.js/blob/dev/src/math/Spherical.js}{spherical} &103000 &35400 &2&448\\
synergy &MIT license &\href{https://github.com/defx/synergy/tree/master/src}{synergy} &310 &3 &5&374\\
t\_test &ISC license &\href{https://github.com/simple-statistics/simple-statistics/blob/main/src/t_test.js}{t\_test} &3400 &226 &6&213\\
validate &MIT license &\href{https://github.com/ehmicky/modern-errors/blob/main/src/subclass/validate.js}{validate} &1500 &18 &2&37\\
zone &MIT license &\href{https://github.com/drawcall/Proton/blob/master/src/zone/Zone.js}{zone} &2400 &274 &3&223\\
\bottomrule
\end{tabular}
% }
\caption{Dataset Details (JavaScript).}
\label{tab: individual_dataset_js}
\end{table*}

%% file: sections/Manual_Protocol.tex
\section{Manual Fixing Protocol}
\label{app:fixing_protocol}

\subsection{Overview}

\textbf{Scope:} Fix only executability errors (prevent test execution) and cascade errors (single root cause affecting multiple tests). Do NOT fix correctness errors, coverage issues, or test logic flaws.

\noindent\textbf{Principle:} Apply minimal necessary changes determinable objectively from error messages and codebase structure.

\subsection{General Decision Framework}

\begin{algorithm}[h]
\caption{Error Fixing Decision Process}
\begin{algorithmic}[1]
\FOR{each error in test execution}
    \STATE identify error type from error message
    \STATE locate matching rule in protocol
    \IF{rule conditions satisfied}
        \STATE apply specified fix
        \STATE re-run tests to verify resolution
        \STATE document: file, line, error type, fix
    \ELSE
        \STATE do not modify
    \ENDIF
\ENDFOR
\end{algorithmic}
\end{algorithm}

% \begin{table}[t]
% \centering
% \small
% \resizebox{\linewidth}{!}{% <-
% \begin{tabular}{ll}
% \hline
% \textbf{Allowed Fixes} & \textbf{Not Allowed} \\
% \hline
% Add missing import & Rewrite test logic \\
% Fix import path (from file structure) & Change assertion values \\
% Add missing parenthesis/bracket & Refactor test structure \\
% Add package declaration (Java) & Add new test cases \\
% Remove hallucinated imports & Rename variables \\
% \hline
% \end{tabular}
% }
% \caption{Minimal Changes Definition}
% \label{tab:minimal_changes}
% \end{table}

\begin{table*}[t]
\centering
\footnotesize
\begin{tabular}{lllc}
\toprule
\textbf{Error Pattern} & \textbf{Condition} & \textbf{Fix Action} & \textbf{LOC} \\
\midrule
\texttt{ModuleNotFoundError: 'X'} & \texttt{\_\_init\_\_.py} exists & Change to \texttt{from pkg.module import X} & 1 \\
\cmidrule{2-4}
 & No \texttt{\_\_init\_\_.py} & Change to \texttt{from module import X} & 1 \\
\cmidrule{2-4}
 & X is external lib & Add \texttt{import X} at top & 1 \\
\midrule
\texttt{ModuleNotFoundError: 'X.Y.Z'} & Extra nesting in path & Remove nesting to match file structure & 1 \\
\midrule
\texttt{ImportError: name 'X'} & Similar name & Fix typo to existing name & 1 \\
\cmidrule{2-4}
 & No similar name & Remove related lines & -- \\
\midrule
\texttt{SyntaxError: EOF/unmatched} & Unbalanced \texttt{()} or \texttt{[]} & Count and add/remove to balance & 1 \\
\bottomrule
\end{tabular}
\caption{Python Executability Error Fixes}
\label{tab:python_exec}
\end{table*}

\begin{table*}[t]
\centering
\footnotesize
\begin{tabular}{lllc}
\toprule
\textbf{Error Pattern} & \textbf{Undefined Name} & \textbf{Fix Action} & \textbf{LOC} \\
\midrule
\multirow{4}{2.5cm}{\texttt{NameError: name 'X' not defined}} & \texttt{np, pd, plt} & Add \texttt{import numpy as np} (etc.) & 1 \\
\cmidrule{2-4}
 & \texttt{unittest, mock} & Add \texttt{import unittest} (etc.) & 1 \\
\cmidrule{2-4}
 & Project class/function & Search project, add import per E1.1 & 1 \\
\midrule
\texttt{FileNotFoundError} & File access error & Fix file access & 1-2 \\
\bottomrule
\end{tabular}
\caption{Python Cascade Error Fixes}
\label{tab:python_cascade}
\end{table*}

\begin{table*}[t]
\centering
\footnotesize
\resizebox{\linewidth}{!}{% <-
\begin{tabular}{lllc}
\toprule
\textbf{Error Pattern} & \textbf{Condition} & \textbf{Fix Action} & \textbf{LOC} \\
\midrule
\texttt{public class should be in file X.java} & Always & Extract package from path, add \texttt{package X;} at line 1 & 1 \\
\midrule
\texttt{cannot find symbol: class X} & Class exists in project & Find file, extract package, add \texttt{import pkg.X;} & 1 \\
\midrule
\texttt{cannot find symbol: method X()} & Missing method import & Add \texttt{import static Y.X;} & 1 \\
\midrule
\texttt{X has private access} & Always & Remove related lines & -- \\
\midrule
\texttt{incompatible types} & Always & Modify to correct types & -- \\
\midrule
Text/comments instead of code & Always & Invalid generation & -- \\
\midrule
\texttt{Mockito exceptions} & Missing import only & Add Mockito imports & 1 \\
\cmidrule{2-4}
 & Wrong mock usage & Remove related lines & -- \\
\bottomrule
\end{tabular}
}
\caption{Java Executability Error Fixes}
\label{tab:java_exec}
\end{table*}

\subsection{Python Fixing Rules}

We categorize fixes into executability errors (preventing test execution) and cascade errors (a single root cause affecting multiple tests).

\subsubsection{Executability Errors}

Table~\ref{tab:python_exec} summarizes fixable executability errors with their conditions and required actions.

\subsubsection{Cascade Errors}

Cascade errors occur when a single error causes multiple test failures. These are mechanically fixable by identifying the undefined name and adding the appropriate import. Table~\ref{tab:python_cascade} lists common patterns.

\subsection{Java Fixing Rules}

Java presents unique challenges due to its stricter syntax requirements and package system. Many executability errors stem from missing package declarations and incorrect import paths.

\subsubsection{Executability Errors}

Table~\ref{tab:java_exec} summarizes Java executability errors. The most distinctive Java error is the missing package declaration, which is deterministically derivable from the file's directory path.

\subsubsection{Cascade Errors}

Java cascade errors primarily involve missing test framework imports and user interaction mocks. Table~\ref{tab:java_cascade} shows fixable patterns. User interaction errors are particularly common in Java due to reliance on \texttt{Scanner} for console input.

\begin{table*}[t]
\centering
\footnotesize
\resizebox{\linewidth}{!}{% <-
\begin{tabular}{lllc}
\toprule
\textbf{Error Pattern} & \textbf{Condition} & \textbf{Fix Action} & \textbf{LOC} \\
\midrule
Test hangs / \texttt{NoSuchElementException} & Scanner input needed & Continue & 1 \\
% \cmidrule{2-4}
%  & Complex interaction & DO NOT FIX & -- \\
\midrule
\multirow{3}{2.5cm}{\texttt{cannot find symbol: @Test, Assert, etc.}} & \texttt{@Test} & Add \texttt{import org.junit.Test;} & 1 \\
\cmidrule{2-4}
 & \texttt{assertEquals} & Add \texttt{import static org.junit.Assert.*;} & 1 \\
\cmidrule{2-4}
 & \texttt{@Mock} & Add \texttt{import org.mockito.Mock;} & 1 \\
\bottomrule
\end{tabular}
}
\caption{Java Cascade Error Fixes}
\label{tab:java_cascade}
\end{table*}

\subsection{JavaScript Fixing Rules}

The primary challenge for JavaScript is calculating correct relative import paths and distinguishing between named and default exports.

\subsubsection{Executability Errors}

Table~\ref{tab:js_exec} presents JavaScript executability errors. Unlike Python's package detection or Java's package declarations, JavaScript requires calculating relative paths between test and source files.

\begin{table*}[t]
\centering
\footnotesize
\resizebox{\linewidth}{!}{% <-
\begin{tabular}{lllc}
\toprule
\textbf{Error Pattern} & \textbf{Condition} & \textbf{Fix Action} & \textbf{LOC} \\
\midrule
\texttt{Cannot find module 'X'} & X is project file & Calculate relative path, fix import & 1 \\
\cmidrule{2-4}
 & X is external & Add \texttt{require('X')} or \texttt{import X} & 1 \\
\midrule
Import/require not working & Check export type & \texttt{export default} → \texttt{require()}, \texttt{export class} → \texttt{\{X\} = require()} & 1 \\
\midrule
\texttt{SyntaxError} & Unbalanced brackets & Count and balance (same as Python) & 1 \\
\cmidrule{2-4}
 & Incomplete generation & Remove incompleted parts & -- \\
\midrule
Empty test suite & Model refused to generate & Invalid generation & -- \\
\bottomrule
\end{tabular}
}
\caption{JavaScript Executability Error Fixes}
\label{tab:js_exec}
\end{table*}

\subsubsection{Cascade Errors}

JavaScript cascade errors involve missing library imports and framework syntax mismatches. Table~\ref{tab:js_cascade} shows common patterns.

\begin{table*}[t]
\centering
\footnotesize
\begin{tabular}{lllc}
\toprule
\textbf{Error Pattern} & \textbf{Undefined Name} & \textbf{Fix Action} & \textbf{LOC} \\
\midrule
\multirow{3}{2.5cm}{\texttt{ReferenceError: X not defined}} & \texttt{expect, assert} & Add \texttt{const \{expect\} = require('chai');} & 1 \\
\cmidrule{2-4}
 & \texttt{THREE, d3} & Add \texttt{const THREE = require('three');} & 1 \\
\cmidrule{2-4}
 & Project component & Calculate path, add \texttt{require('../X')} & 1 \\
\midrule
Wrong framework syntax & Jest in Mocha, etc. & Convert to correct framework & 1-2 \\
% \cmidrule{2-4}
%  & Ambiguous case & DO NOT FIX & -- \\
\bottomrule
\end{tabular}
\caption{JavaScript Cascade Error Fixes}
\label{tab:js_cascade}
\end{table*}

\subsection{Validation Protocol}

Each fix must pass a five-step validation process to ensure objectivity and reproducibility, as shown in Table~\ref{tab:validation}. This protocol is applied identically by all annotators to minimize subjectivity.

\begin{table*}[t]
\centering
\small
% \resizebox{\linewidth}{!}{% <-
\begin{tabular}{cll}
\toprule
\textbf{Step} & \textbf{Check} & \textbf{Action} \\
\midrule
1 & Error resolved? & Re-run test, verify specific error gone \\
2 & New errors? & Treat as separate issues, apply protocol \\
3 & Changes minimal? & Must directly address error, nothing extra \\
4 & Deterministic? & Another annotator would make same fix \\
5 & Documented? & Log: file, line, error type, rule, fix, LOC \\
\bottomrule
\end{tabular}
% }
\caption{Per-Fix Validation Checklist}
\label{tab:validation}
\end{table*}